\documentclass[%
 reprint,
 amsmath,amssymb,
 aps,
]{revtex4-1}

\usepackage{float}
\usepackage{graphicx}
\usepackage{dcolumn}
\usepackage{bm}

\usepackage{xcolor}
\begin{document}

\preprint{APS/123-QED}

\title{Ultrahigh molecular recognition specificity of competing DNA oligonucleotide 
strands in thermal equilibrium: a cooperative transition to order}

\author{Marc Schenkelberger}
\author{Christian Trapp}%
 \author{Timo Mai} 
 \altaffiliation[formerly at]{Experimental Physics I, Bayreuth University, 
94550 Bayreuth, Germany\\}
 \author{Mina Mohammadi-Kambs}
 \author{Varun Giri}  
 \altaffiliation[presently at]{BASF Ludwigshafen\\}
 \author{Albrecht Ott}
\affiliation{%
 Biological Experimental Physics, Saarland University, 66123 Saarbr\"ucken, Germany\\
}%




\date{\today}

\begin{abstract}
The specificity of molecular recognition is important for molecular self-organization. 
A prominent example is the biological cell where a myriad of different molecular receptor pairs recognize their binding 
partner with astonishing accuracy within a highly crowded molecular 
environment. In thermal equilibrium it is usually admitted 
that the affinity of recognizer pairs only depends on the nature of the two binding 
molecules. Accordingly Boltzmann factors of binding energy differences relate 
the molecular affinities among different target molecules that compete for the 
same probe. Here, we consider the molecular recognition of short DNA oligonucleotide 
single strands. We show that a better matching oligonucleotide can prevail against a disproportionally 
more concentrated competitor with reduced affinity due to a mismatch. We investigate the situation using fluorescence based techniques, among them F\"orster Resonance Energy Transfer (FRET) and Total Internal Reflection 
Fluorescence excitation (TIRF). 
We find that the affinity of certain strands appears considerably reduced only as long as a better matching competitor is present. Compared to the simple Boltzmann picture above we observe increased specificity, up to several orders of magnitude. 
We interpret our observations based on an energy-barrier of entropic origin that occurs if two competing oligonucleotide strands occupy the same probe simultaneously. Due to their differences in binding microstate distributions, the barrier affects the binding affinities of the competitors differently. Based on a mean field description, we derive a resulting expression for the free energy landscape, a formal analogue to a Landau description of phase transitions reproducing the observations in quantitative agreement as a result of a cooperative transition. The advantage of improved molecular recognition comes at no energetic cost other than the design of the molecular ensemble and the presence of the competitor. As a possible application, binding assays for the detection of single nucleotide polymorphisms in DNA strands could be improved by adding competing strands. It will be interesting to see if mechanisms along similar lines as exposed here, contribute to the molecular synergy that occurs in biological systems. Macromolecular building blocks that mutually proofread their binding states to cooperatively improve the precision of self-assembly offer a clear advantage for the design of complex, self-organizing structures.

\begin{description}
\item[Usage]
Secondary publications and information retrieval purposes.
\item[PACS numbers]
May be entered using the \verb+\pacs{#1}+ command.
\end{description}
\end{abstract}

\pacs{Valid PACS appear here}

\maketitle


\section{\label{Introduction}Introduction}

Many chemical reactions in biochemistry, molecular medicine, or biotechnology 
rely on the specificity of molecular recognition. Specific binding is crucial to 
the formation of dedicated macromolecular complexes \cite{Lehn:1990}. Molecular specificity plays an important role in biology, driving catalysis, information transmission, and molecular self-organization. Accordingly there has been a long-standing interest in molecular binding. The well-known 
`Lock and Key' model explains the recognition specificity of enzymes solely based 
on their molecular shape \cite{Fischer:1894}. The `Induced Fit' and the `Conformational Proofreading' 
models \cite{Savir:2007} go beyond that. They show that the deformation 
of a (flexible) molecule upon binding can lead to an enthalpic barrier that increases 
binding specificity at the expense of binding affinity. Cooperative binding through multiple sites can increase binding affinity and specificity 
at the same time. DNA hybridization is an example: binding of one or a few complementary 
base pairs increases the binding probability of other complementary bases from the same strand, 
and this increases recognition specificity. Cooperative binding of several molecules 
to the same DNA strand is observed for instance in transcription regulation. This can increase specificity 
along similar lines\cite{Hippel:1982}. The `Pre-existing Equilibrium' or `Monod, Wyman, and 
Changeux' model \cite{Monod:1965} considers allosteric molecules that possess several conformational 
states with different affinities. The state of increased affinity to a ligand will 
be stabilized as long as a third, different molecule remains attached to the allosteric 
site.

In the examples above specificity solely depends on the design of the molecular 
recognisers. A different case is the use of energy to create thermal non-equilibrium 
and increase specificity. In their seminal papers \cite{Hopfield:1974, Ninio:1975} Hopfield and Nin${\tilde \i}$o discussed how DNA polymerases consume ATP (adenosine triphosphate) as energy source to reduce 
error during the incorporation of single bases in a DNA copying process. Due to 
the laws of thermodynamics, in competition and in thermodynamic equilibrium (without 
energy source), the error is fixed by the ratio of the individual binding constants 
of the incorrect and correct bases, \textit{K}\textsubscript{incorr }/ \textit{K}\textsubscript{corr}, 
(in vicinity of 10\textsuperscript{-1} or 10\textsuperscript{-2}) and the law of mass action. Combining 
several energetically excited binding states, which are selective 
and proceed one after another, polymerases reduce error by several orders 
of magnitude at the expense of both, energy and biochemical side reactions. The 
mechanism has since become the paradigm of `kinetic proofreading'.

There is no `kinetic proofreading' process for DNA hybridization, and non-equilibrium 
states, often metastable, are considered detrimental to the accuracy of the recognition 
process \cite{Bishop:2006}.  Accordingly, it is generally believed that the molecular recognition among competing DNA strands performs best in thermal equilibrium where interactions can be neglected \cite{Bhanot:2003, Bishop:2006,Zhang:2012}. The lower limit of the recognition error is then given by the ratio of the individual binding constants of the competing strands as obtained from pairwise assessments.

DNA duplex stability arises from hydrogen bonding and base stacking interactions that encompass van der Waals -, electrostatic - and hydrophobic interactions between adjacent base pairs. The well-established nearest-neighbor model considers that the energy of a nucleic acid duplex is the sum of these nearest-neighbor (NN) interactions \cite{devoe62, borer74, breslauer86}. The empirical nearest neighbor helix propagation parameters (one for each of the 10 possible base-pair doublets) serve to predict the binding free energy of a particular duplex sequence. Other parameters provide corrections for duplex initiation, AT terminal pairs, or a symmetry penalty in case of self-complementary sequences. The NN model adequately predicts oligonucleotide duplex melting temperatures in bulk solution \cite{santalucia04}. Elaborate datasets of Watson-Crick NN parameters \cite{SantaLucia:1998} provide the basis for nucleic acid secondary structure - or melting temperature prediction software as for instance the DINAMelt web server \cite{Markham:2005}, Mfold \cite{zuker03}. Nupack \cite{zadeh:2011} is one of the latest packages for quantitative prediction of DNA hybridization free energy in thermal equilibrium. Since these software packages are semi-empirical, scope and precision of the underlying data do matter. If conditions are changed, the accuracy of predictions is reduced  \cite{srinivas13, zhang09, zhang12, idili17, moreira2005, wolk15}. Improvements to simple theoretical models and empirical parameters remain under discussion \cite{Hooyberghs:2009, Huguet:2010, Hadiwikarta:2012, Vologodskii18}. Other binding contributions besides nearest neighbor contributions have been shown to make non-negligible contributions in certain settings \cite{Mohammadi-Kambs:2019}.

One of the most challenging aspects of oligonucleotide technology is the recognition 
of perfectly matching molecules in an environment of many similar competitors. 
This is particularly difficult if the oligonucleotides differ by as little as a 
single base, as for instance in the search for single nucleotide polymorphisms of medical importance \cite{Lee:2014}. Although the NN model predicts a loss of a factor of 1000 in binding affinity for oligonucleotides similar to what we address in this study \cite{Mohammadi-Kambs:2019}, all experimental studies we became aware of yielded much smaller differences. In pairwise assessments the affinity  weakened only by about a factor of 10 compared to the perfect match \cite{Bonnet:1999, Altan-Bonnet:2012, zhang12, Wang:2015} and references therein. In previous work we showed that the deviation from the NN model is due to binding states other than the double helix \cite{Mohammadi-Kambs:2019}. 

At the same time, DNA hybridization can perform surprisingly well in a highly competitive environment. An example is in situ hybridization where a nucleotide strand of given sequence, up to a few hundred base pairs long, exclusively binds  to its complement in the crowded molecular environment of a biological cell \cite{Brecevic:2006}. 

Here our aim is to investigate the molecular recognition specificity of oligonucleotide 
hybridization in a competitive situation \cite{Zhang:2005, Horne:2006, Bishop:2008, Cherepinsky:2010, Williams:2011, Altan-Bonnet:2012}. Others observed unexpected behavior in such settings \cite{Zhang:2005, Williams:2011}. Previously we showed that a competitive situation may well entail cooperativity  if competitors overlap on the binding site \cite{Mohammadi-Kambs:2019}, modifying the ratio of their binding constants compared to the pairwise assessment. In the following we investigate the situation of  oligonucleotides that differ only in a single base, competing for the same probe sequence. We show that binding can be much more specific than expected from pairwise considerations. We propose a mechanism that increases specificity due to interactions among the target strands, yielding a collective transition towards more accurate probe occupancy. The proposed cooperative mechanism quantitatively reproduces our observations. It is robust with respect to the details of the model.

\section{\label{Material and Methods}Material and Methods}

\begin{figure*}[!htb]
\includegraphics[width=500pt, height=431pt, keepaspectratio=true]{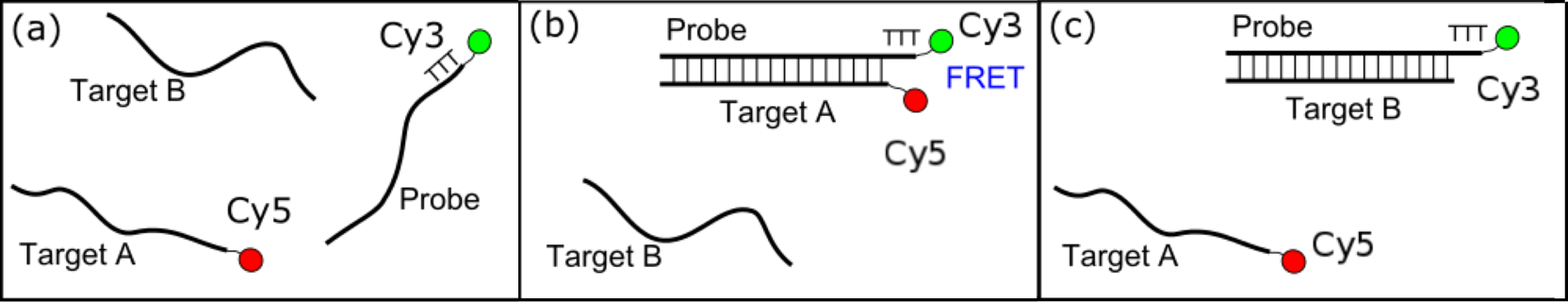}
\caption{FRET (F\"orster Resonance Energy Transfer)\textbf{(a)} Target A is labeled with the fluorescent dye Cy5 (red dot) at the 5' end of the strand while the probe is labeled with Cy3 (green dot) 
at the 3' end. Target B remains unlabeled. \textbf{(b)} Upon hybridization of target 
A to the probe Cy3 emission is quenched by Cy5, and Cy5 will start to 
emit although only Cy3 is excited. A spacer of 3 Thymine bases is required to avoid contact quenching. \textbf{(c) }Hybridization of the unlabeled target B to the probe does not quench Cy3 emission: Cy5 does 
not emit. In order to detect the hybridization of target B in the presence of target 
A, labeled and unlabeled targets need to be interchanged (not shown).}
\label{fig:FRETfig}
\end{figure*}

\subsection{\label{Oligo}Oligodeoxyribonucleotide sequences and buffer solutions}

All oligonucleotides are from Metabion (Martinsried, Germany), HPLC 
purified. The concentrations of oligonucleotides are checked by UV absorption at 
260 nm. We choose sequences at random, unless stated otherwise, such that no stable 
secondary structures are predicted under the experimental conditions by the DINAMelt 
web server \cite{Markham:2005}. We use SSC and SSPE buffer for hybridization prepared 
with deionized and purified water (Sartorius, Arium Advance); pH is adjusted to 7.5. 
Experiments are performed with 3xSSC 
(TIRF (total internal reflection fluorescence) and FRET (Fo\"rster resonant energy transfert), 0.57 M monovalent ions) or 5xSSPE (DNA microarray, 0.95 M monovalent ions). Theses buffers 
 are often employed in DNA microarray studies. High salt concentrations screen 
unwanted charge interactions for example with the surfaces. Moreover high salt 
enables fast and accurate DNA hybridization at low oligonucleotide concentrations. 
Under these conditions we previously showed that for pairwise recognition (without competition) a theoretical 
equilibrium description works very well. With the 
here employed parameter set we predicted the impact of a single mismatch on binding to microarray surfaces with so far unreached accuracy \cite{Naiser2009}. In part this was due to surface grafting via dendrimers \cite{Trevisol:2003} that produced data closely matching bulk hybridization.

\subsection{\label{Coating}Dendrimer coating of the glass slides used in surface based measurements}

We use standard microscopy coverslips with a diameter of 20 mm as substrates for 
DNA immobilization. Coverslips are cleaned with Deconex (Borer Chemie, Switzerland) 
and rinsed with purified water. Silanization and functionalization with dendrimeric 
molecules (Cyclotriphosphazene PMMH, Generation 2.5 or 4.5, Aldrich) bearing aldehyde 
endgroups for immobilization of amino-modified DNA is performed according to \cite{Trevisol:2003}. 
The functionalized surface consists of a roughly one micrometer thick coating of low 
density dendrimers \cite{Naiser:2008}.

\subsection{\label{Grafting}DNA grafting to dendrimer coated slides}

For immobilization of DNA probes we follow the protocol in \cite{Trevisol:2003} with the following modifications: Immobilization time is reduced to one hour. 
We limit sodium borohydride treatment to 15 min and add 25 percent (v/v) EtOH to 
reduce the formation of hydrogen on the substrate and avoid inhomogeneities in 
grafting density. This modified protocol does not reduce the number of surface 
bound DNA molecules.

\subsection{\label{FRET}F\"orster Resonance Energy Transfer (FRET)}

We employ F\"orster Resonance Energy Transfer (FRET) to assess the hybridization 
of target and probe in bulk. One of the two competing target molecules is labeled 
with the acceptor molecule Cy5 at the 5' end. The complementary probe molecule is 
labeled with the donor molecule Cy3 at the 3' end. The other competing target 
molecule is not labeled. To avoid contact quenching we set the distance between 
the fluorescent dyes Cy3 and Cy5 by adding three T bases 
to the 3' end of the probe molecule. The three different oligonucleotides are mixed 
in a standard reaction tube, incubated at 44 $^{\circ}$C. Immediately after mixing, 200 
\ensuremath{\mu}l of the solution is transferred into a non-absorptive 96 well 
microplate (Nunc, Germany). The fluorescent signal is immediately determined using 
a plate-reader (Polar Star Optima, BMG Labtech, Germany). We use 544 nm as the 
excitation wavelength of Cy3 and measure either the emission of Cy3 at 580 nm (donor 
channel) or of Cy5 at 670 nm (acceptor channel). See figure~\ref{fig:FRETfig} for illustration of the working principle.

\subsection{\label{TIRF}Total Internal Reflection Fluorescence (TIRF) for time dependent observation 
of surface based DNA hybridization}

\begin{figure}[!h]
\includegraphics[width=225pt, height=156pt, keepaspectratio=true]{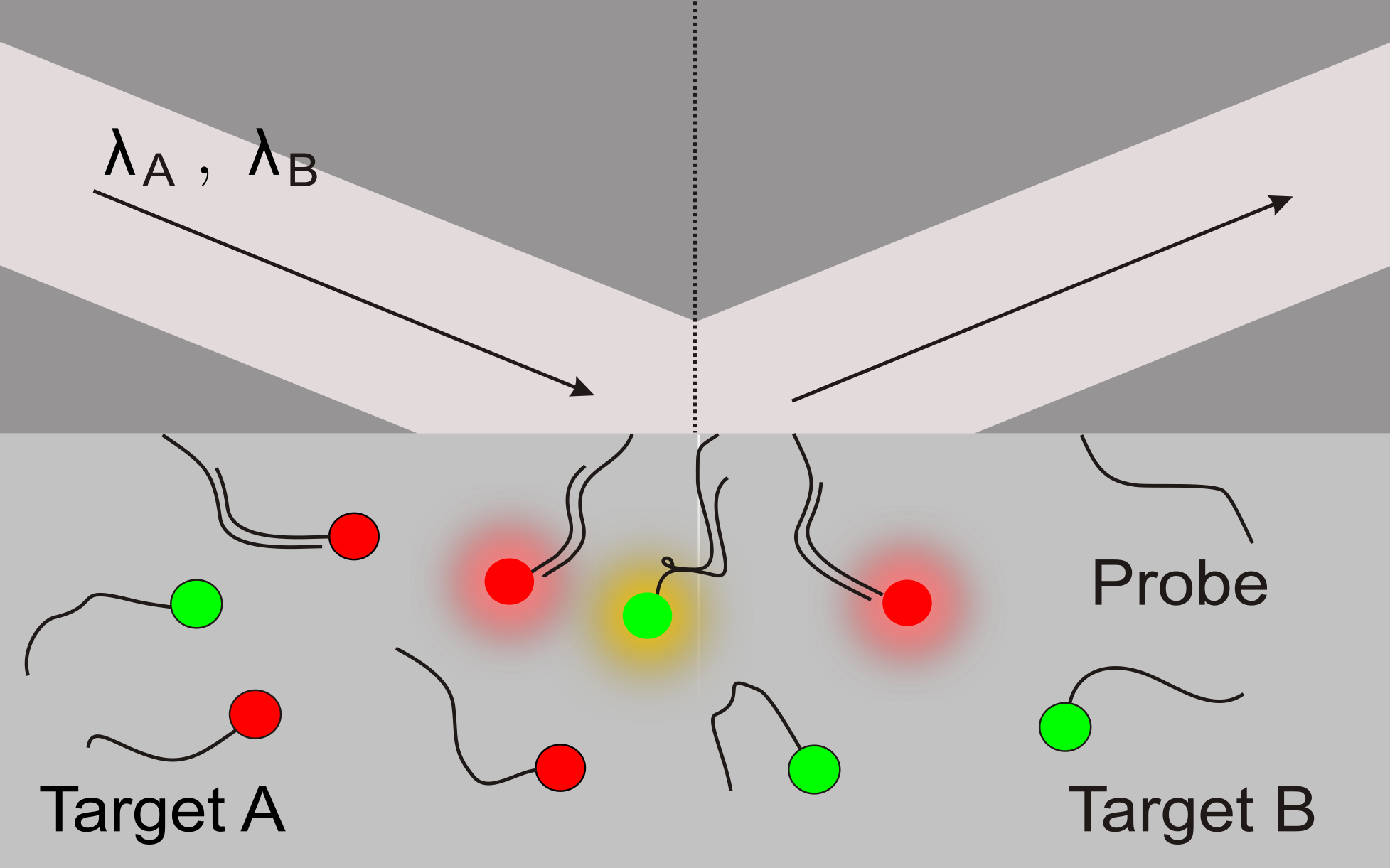}
\caption{\textbf{TIRF (Total Internal Reflection Fluorescence)} \textbf{detection.} 
Competing target species A, Cy5 (red) labeled, and B, Cy3 (green) labeled, bind 
to surface immobilized probe molecules. Two 
 evanescent fields of wavelengths $\lambda _{A}  $  and $\lambda _{B}  $  
excite the fluorescent dyes of the competing targets up to about 
100 nm away from the surface (excitation is indicated by aureoles). The time course of the intensity of the emitted fluorescent light is recorded by two photomultiplier detectors (one for each wavelength). 
}
\label{fig:TIRFfig}
\end{figure}

TIRF relies on the evanescent field penetrating into the medium of lower refractive 
index at the point of total reflection by a distance of typically 100 nm (Fig.~\ref{fig:TIRFfig}). This field excites fluorescent dyes at the interface. 

A cover glass with immobilized DNA is fixed as part of the observation chamber mounted on the \textit{xy}-stage 
of an inverted microscope (Axiovert 135, Zeiss, Germany). The excitation beam reaches 
the cover glass through a dove prism and a layer of immersion oil (see supplementary 
material S1 for a scheme of the optical path). To enable simultaneous detection 
of Cy3 and Cy5 labeled molecules during competitive experiments, two 
lasers of different wavelengths (DiodePumpedSolidState, 532 nm and HeNe, 633 nm) excite the fluorescent 
dyes. Before entering the prism, laser beams are chopped, expanded 
and focused onto the observation chamber. The excitation power (44 \ensuremath{\mu}W) 
is as low as possible to minimize photo bleaching. Fluorescence emission is collected 
through a microscope objective (Achromat 10x, 0,25 N.A. Zeiss, Germany). A beam 
splitter directs the emitted light through the emission filters (Cy3 channel: HQ585/40, 
Cy5 channel: HQ680/30, Chroma Technology, USA), each followed by a photomultiplier 
(H9305-04, Hamamatsu). Lock-in amplified signals are recorded using a PC. A PID 
control equipped with a PT100 sensor controls the temperature of the experiment. 
To avoid a temperature drop towards the observation window, we apply an electrical 
heating current through its Indium tin oxide (ITO) coating. The observation chamber 
is of circular cross section (viewed from top) with a diameter of 4\textit{ }mm 
and a height of 2.5\textit{ }mm. It is filled via openings on its sides before 
the start of an experiment. We reuse the same coverslips with grafted DNA for several 
hybridization experiments. Removal of hybridized DNA for the following
experiment is performed by treatment with 10\textit{mM} NaOH for 1 minute at 44 $^{\circ}$C. 
In case of PM and MM1 (table ~\ref{tab:tableone}) exchange of the dyes between the 
two did not lead to any detectable difference in PM hybridization. For all experiments the green channel is used for monitoring the oligonucleotide of interest, since the photomultipliers respond more efficiently to this wavelength, and the temporal resolution is higher.

\begin{table*}
\caption{\label{tab:tableone} Sequences and experimentally determined binding affinities of 
all target molecules under study. The perfect matches (PM, PM*) are complementary 
to the probe sequence. The (mismatched) targets MM1 and MM2 possess a single non-complementary 
base, targets MM3, MM4, MM* possess two of them. The mismatch in MM* is degenerate, 
and this may well explain that the binding affinity is close to PM* \cite{Naiser:2009}. With the type of surfaces used here, immobilisation of the probes does not lead to significant modification of the 
relative binding affinities compared to bulk \cite{Naiser:2009}}
\begin{ruledtabular}
\begin{tabular}{cccc}
\textrm{Target}&\textrm{Sequence}&\multicolumn{2}{c}{Binding Constant 10-7 1/M}\\
 &&surface&bulk\\ \hline
PM&AAG-GAT-CAG-ATC-GTA-A&$90 \pm 50$&$20 \pm 5$\\
MM1&AAG-GAT-CA\textcolor{red}C-ATC-GTA-A&$6 \pm 1$&$1.3 \pm 0.5$\\
MM2&AAG-GAT-CAG-ATC-G\textcolor{red}CA-A&$40 \pm 40$&$20 \pm 4$ \\
MM3&AAG-GAT-C\textcolor{red}{TC}-ATC-GTA-A&$1 \pm 0.1$&-\\
MM4&AAA-GAT-CAG-ATC-G\textcolor{red}AA-A&$10 \pm 3$&-\\
PM*&GGG-CAG-CAA-TAG-TAC&$200 \pm 30$&-\\
MM*&GGG-CAG-C\textcolor{red}{TT}-TAG-TAC&$200 \pm 80$&-\\
\end{tabular}
\end{ruledtabular}
\end{table*}

\subsection{\label{Microarrays}DNA microarrays}

DNA microarrays are synthesized in situ using a maskless photolithographic technique 
based on NPPOC phosphoramidites (details in \cite{Naiser:2009} and references therein). Probes 
are surface bound, however, DNA is polymerized 
`from' the surface (instead of grafted `to' ), leading to a better control of 
surface deposition and density \cite{Michel:2007}. A further advantage of microarrays is that different strands 
can be arranged in different locations, which enables us to perform experiments in 
parallel. However, the optical control of the photolithographic process produces 
a higher number of sequence errors compared to standard nucleotide synthesis \cite{Naiser:2009}.
After adding fluorescently labeled target molecules to the surface bound probes 
and waiting for equilibrium, we quantitatively acquire the fluorescence intensity 
of the hybridized DNA by imaging the microarray surface through an epifluorescence 
microscope.

\begin{figure}[!htb]
\includegraphics[width=280pt, height=270pt, keepaspectratio=true]{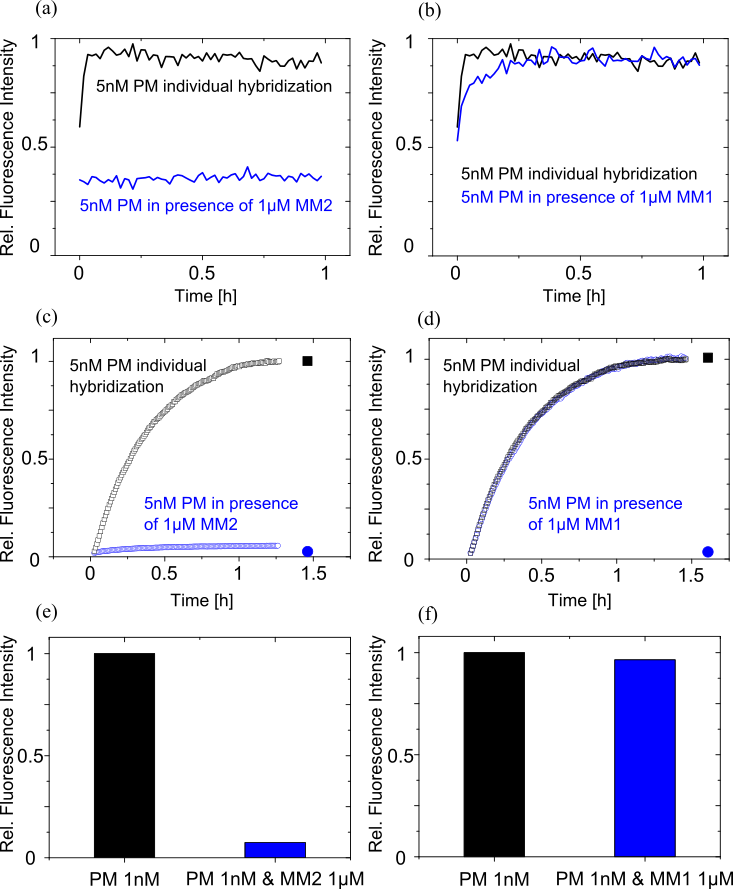}
\caption{ \textbf{Standard (left) vs. highly specific binding (right) competitive DNA hybridization.} In all graphs we show the hybridization of the target PM to its complementary probe (table ~\ref{tab:tableone} for sequences). PM hybridization in the presence of a competitor is in blue, without competitor is in black. Fluorescence intensity (vertical axes) is proportional to the amount of probe bound PM. The scale is given as relative to the equilibrium value in the absence of competing targets. The competing targets are either MM1 or MM2, neither labelled in the experiment nor shown in the graphs. MM1 differs from PM by a mismatch in the middle of the strand, MM2  towards one end. The initial PM concentration is 5 nM (1 nM in (e), (f)). Temperature is 44$^{\circ}$C.  \textbf{(a), (c), (e), `PM vs. MM2'} leads to standard specificity: the degree of PM probe occupancy is strongly affected by the presence of 1\ensuremath{\mu}M MM2. \textbf{(a)}, PM binding in bulk as a function of time, observed in FRET. \textbf{(c)}, PM binding to surface bound probes as a function of time, observed using TIRF. For comparison, filled blue circles correspond to the equilibrium duplex concentration as predicted by a Boltzmann description (Eq.~\eqref{eqone}) using the binding constants from table ~\ref{tab:tableone}. \textbf{(e)}, Binding to a DNA microarray feature (same sequence as probe), observed using epifluorescence microscopy (endpoint values in equilibrium). \textbf{(b), (d), (f), `PM vs. MM1'} exhibits high specificity. Otherwise than the choice of the competing target (MM1 instead of MM2) experimental conditions are identical to (a), (c), (e).  The presence of 1\ensuremath{\mu}M MM1 does not affect the hybridization of the 5nM or 1 nM PM in any detectable way, revealing orders of magnitude higher specificity of PM hybridization than in (a), (c), and (e). 
}
\label{fig:SBHSB}
\end{figure}

\section{\label{Results}Results}

In a first set of experiments we consider a DNA-strand of 16 base pairs, the probe. 
Its sequence was chosen at random given the constraint that the Dinamelt server \cite{Markham:2005} does not predict any significant secondary structures. 
We consider three different target strands: PM is a perfectly matching complement 
to the probe. MM1 and MM2 differ from the PM in only 
a single, non-matching base (table ~\ref{tab:tableone}). To monitor the amount of hybridized 
PM in the presence of a competing strand, we use F\"orster Resonance Energy 
Transfer (FRET) in bulk (Fig.~\ref{fig:FRETfig}). We use TIRF (Total Internal Reflection Fluorescence) detection in the context of surface grafted probes (Fig.~\ref{fig:TIRFfig}). We find that in the 
case where PM and MM2 are present simultaneously and compete for the probe (Fig.~\ref{fig:SBHSB}(a, c, e)), high concentrations of MM2 reduce the number of PM occupied probes 
substantially. However, in the case `PM vs. MM1' (Fig.~\ref{fig:SBHSB}(b, d, f)), an equivalent 
excess of MM1 targets over PM targets does not lead to any detectable decrease 
of the PM hybridization level.

If targets A and B compete for probe molecules to form a duplex, one expects the 
ratio of equilibrium duplex concentrations ${\lbrack D^{A} ]_{eq} }/{\lbrack D^{B} ]_{eq} }  $  
to depend on the ratio of the equilibrium target concentrations ${\lbrack A]_{eq} }/{\lbrack B]_{eq} }  $, 
and the ratio of the individual binding affinities $K_{A}$ and $K_{B}$ as:
\begin{equation}
 \frac{\lbrack D^{A} ]_{eq} }{\lbrack D^{B} ]_{eq}} = \frac{\lbrack A]_{eq}   K_{A} }{\lbrack B]_{eq}   K_{B}}= \frac{\lbrack A]_{eq}}{\lbrack B]_{eq}}\exp(\frac{\Delta \Delta G}{k_{B}T})
 \label{eqone}.
 \end{equation}
 
where $\Delta \Delta G=\Delta G^{A} -\Delta G^{B}  $  is the difference between 
the individual effective binding free energies of the competing strands, $\Delta G^{A}  $  
and $\Delta G^{B}  $, $k_{B}  $  is the Boltzmann constant, $N_{A}  $  is the 
Avogadro number and \textit{T} the temperature.

We determine the individual binding affinities of the targets $K^{PM}  $ and $K^{MM1}  $ 
 in experiments without competition. We use an extended Langmuir-type adsorption isotherm [23], taking into account the reduction of free target strands in bulk due to hybridization (supplementary material 
S2), to determine the binding affinities. 
We find that the binding affinities $K^{PM} $  and $K^{MM1} $  differ by about one order of magnitude (table ~\ref{tab:tableone}, see supplementary material S2,
S3.1, S3.2, S4.1 and S4.2 for data and derivation of the binding constants). However, in the 
case `PM vs. MM1' the PM hybridizes as if the highly concentrated competitor 
MM1 was not present. The experimentally observed mismatch discrimination 
is improved over the predictions of Eq.~\eqref{eqone} by several orders of magnitude.

\begin{figure}[htb!]
\includegraphics[width=240pt, height=472pt, keepaspectratio=true]{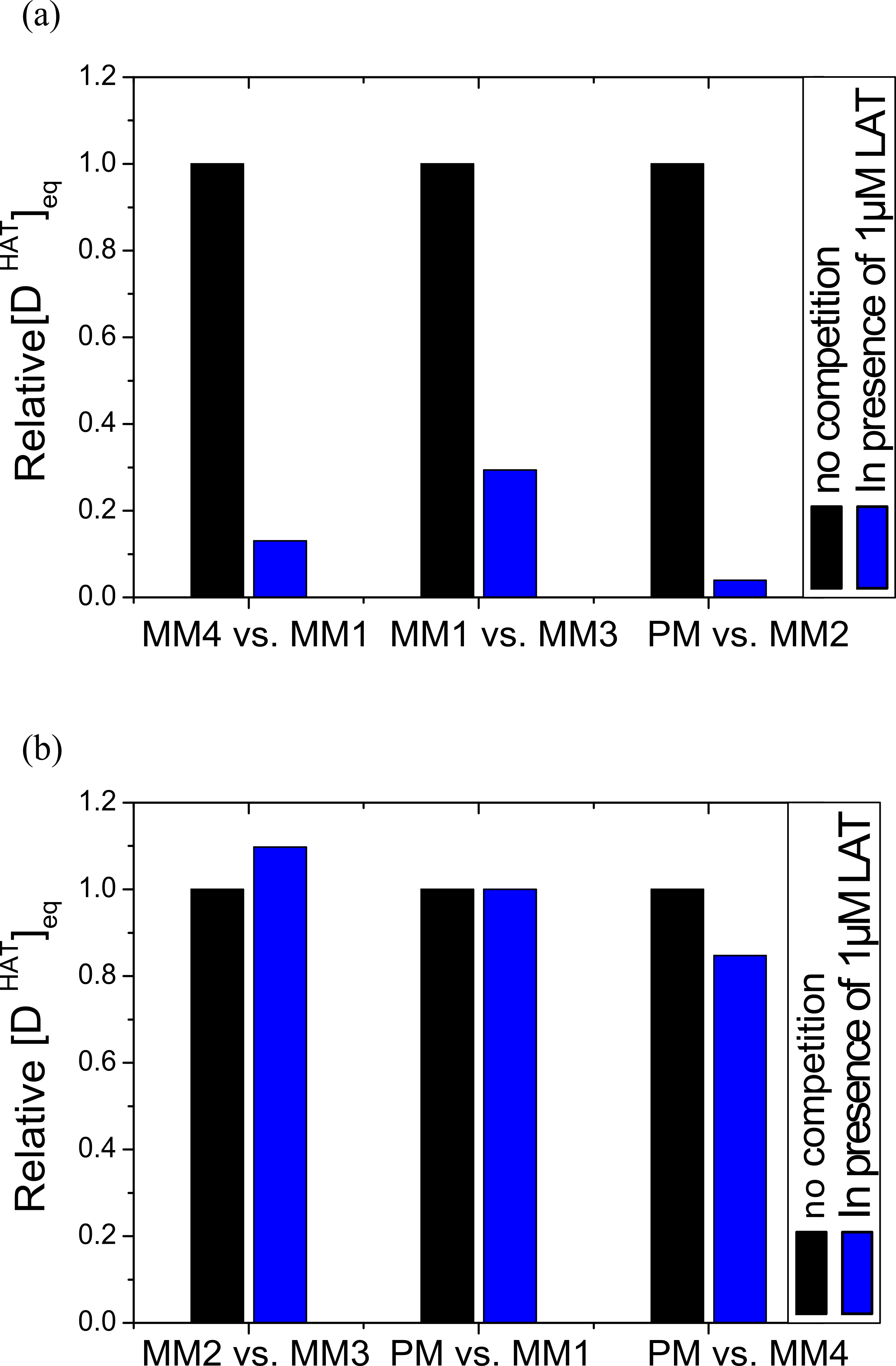}
\caption{\textbf{Competitive DNA hybridization among different sequences}\\ \textbf{(a), (b),} Equilibrium PM probe occupancy (at 44$^{\circ}.$C) as observed by TIRF for several pairs of competing targets chosen from table ~\ref{tab:tableone} : For each pair of competing targets we monitor the equilibrium probe occupancy of the HAT (the high affinity target), $\lbrack D^{HAT} ]_{eq}  $, only. The LAT (low affinity target) occupancy is not shown. The HAT occupancy of the probe strand is given without (black) and in the presence (blue) of the LAT. $\lbrack D^{HAT} ]_{eq}  $ is scaled relative to its equilibrium value in the absence of any competing strands. The LAT concentration is 1 \ensuremath{\mu}M; the HAT concentrations are 10 nM (`MM4 vs. MM1' and `MM1 vs. MM3') or 5 nM (all others). As in Fig.~\ref{fig:SBHSB} we clearly distinguish (a), standard specificity following Eq.~\ref{eqone} and (b), high specificity of the HAT where the presence of the competitor does not show in spite of elevated LAT concentrations.}
 \label{fig:Diffsequences}
\end{figure}

Competitive hybridization experiments among other strands related to the PM sequence, 
including target molecules MM2, MM3 and MM4 (differing from PM in two non-matching base 
pairs, table ~\ref{tab:tableone}), reveal two characteristic scenarios: either standard specific 
systems (Fig. ~\ref{fig:Diffsequences}(a)), where the presence of a `Low Affinity Target' (LAT) reduces 
the equilibrium duplex concentration of a `High Affinity Target' (HAT) as predicted 
by Eq.~(\eqref{eqone}), or highly specific systems (Fig.  ~\ref{fig:Diffsequences}(b)) where the HAT hybridizes as if 
the LAT was not present (supplementary material S4.1 and S4.2 for experimental data).

We checked for equilibrium conditions in the case of PM vs. MM1 (supplementary 
material S5). The observed behavior is perfectly consistent with respect to perturbations in temperature. The observations are independent of the order of sample introduction and do not change during extended time periods.
We verify that the sum of the red and green emission intensity in equilibrium corresponds to complete single occupation of all surface bound probes (see supplementary material S6). 
To test if our result is sequence related, we randomly choose a second 
motif in related literature [15]. After shortening to adjust the denaturation temperature 
for our experimental setup, the investigation of the sequences PM* and MM* (Table 
1) again yields a strong deviation from Eq.~\eqref{eqone}  (supplementary material S7). 

\begin{figure}[]
\includegraphics[width=225pt, height=472pt, keepaspectratio=true]{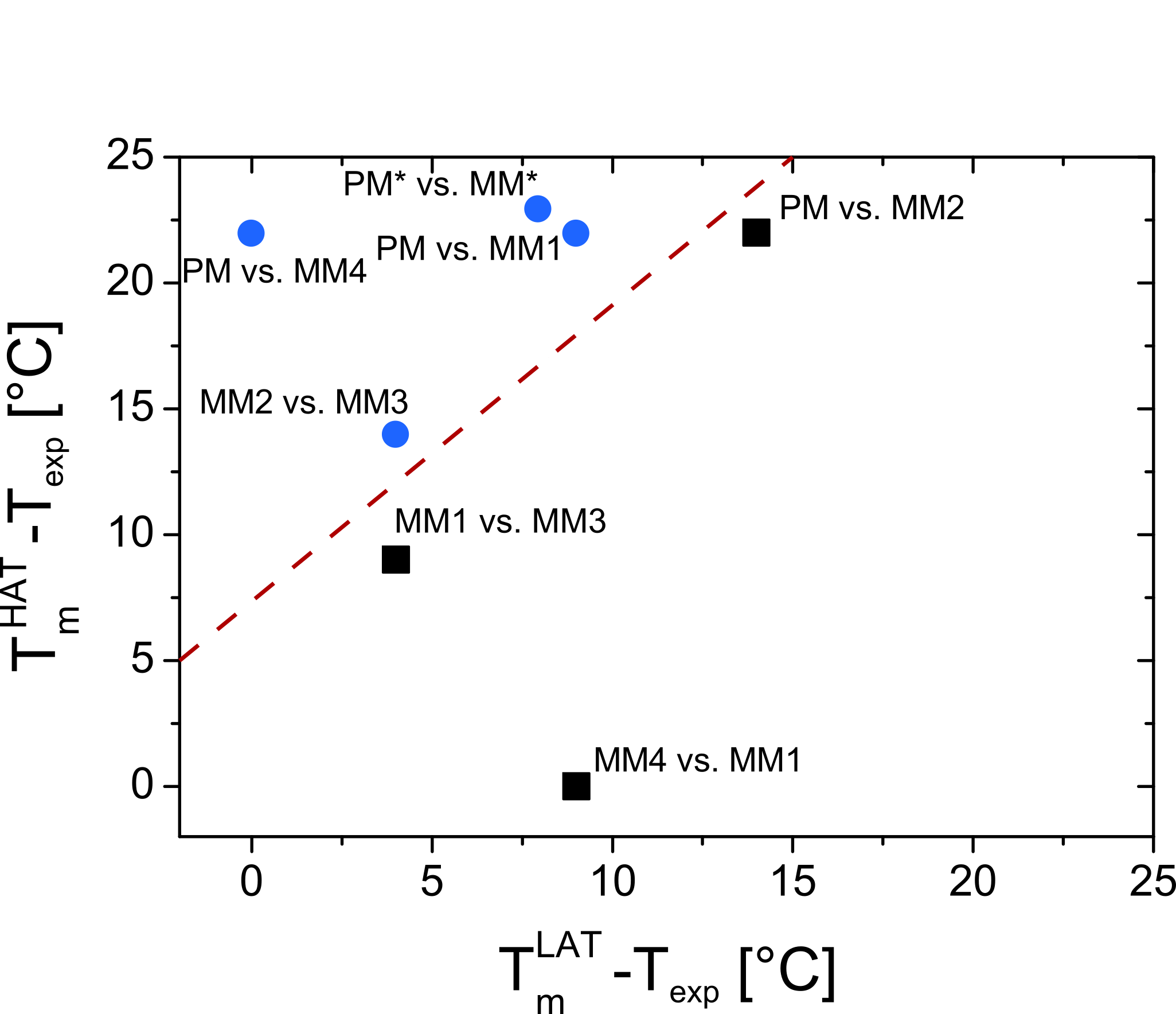}
\caption{\textbf{Comparison of melting temperature differences among competing strands.}\\ \textbf{(c)}, Melting temperatures as predicted by the Dinamelt web server (see table ~\ref{tab:tableone} for sequences) for studied HAT/LAT pairs. Blue: highly specific cases, black: standard specific cases, $T_{m}^{HAT}  $, $T_{m}^{LAT}  $  are the melting temperatures of the two competing targets,  $T_{exp}  $ = 44$^{\circ}$C denotes the temperature of the experiment. 
 }
 \label{fig:MeltingTemperatures}
\end{figure}

Looking at our entire data set, we note that we experimentally observe standard 
specificity if the predicted melting temperature from the Dynamelt web server $T_{}$  of the HAT and the LAT are comparable (Fig. ~\ref{fig:MeltingTemperatures}) while increased specificity occurs in cases where the melting temperatures of the two competitors differ by a certain amount, roughly 10 $^{\circ}$C (supplementary 
material S8).

We observe that the degree of deviation from Eq.~\ref{eqone}  diminishes with lower  
temperature of the experiment, but the deviation does not disappear (supplementary 
material S9). This establishes entropy as a driving force. 

\begin{figure}[]
\includegraphics[width=225pt, height=472pt, keepaspectratio=true]{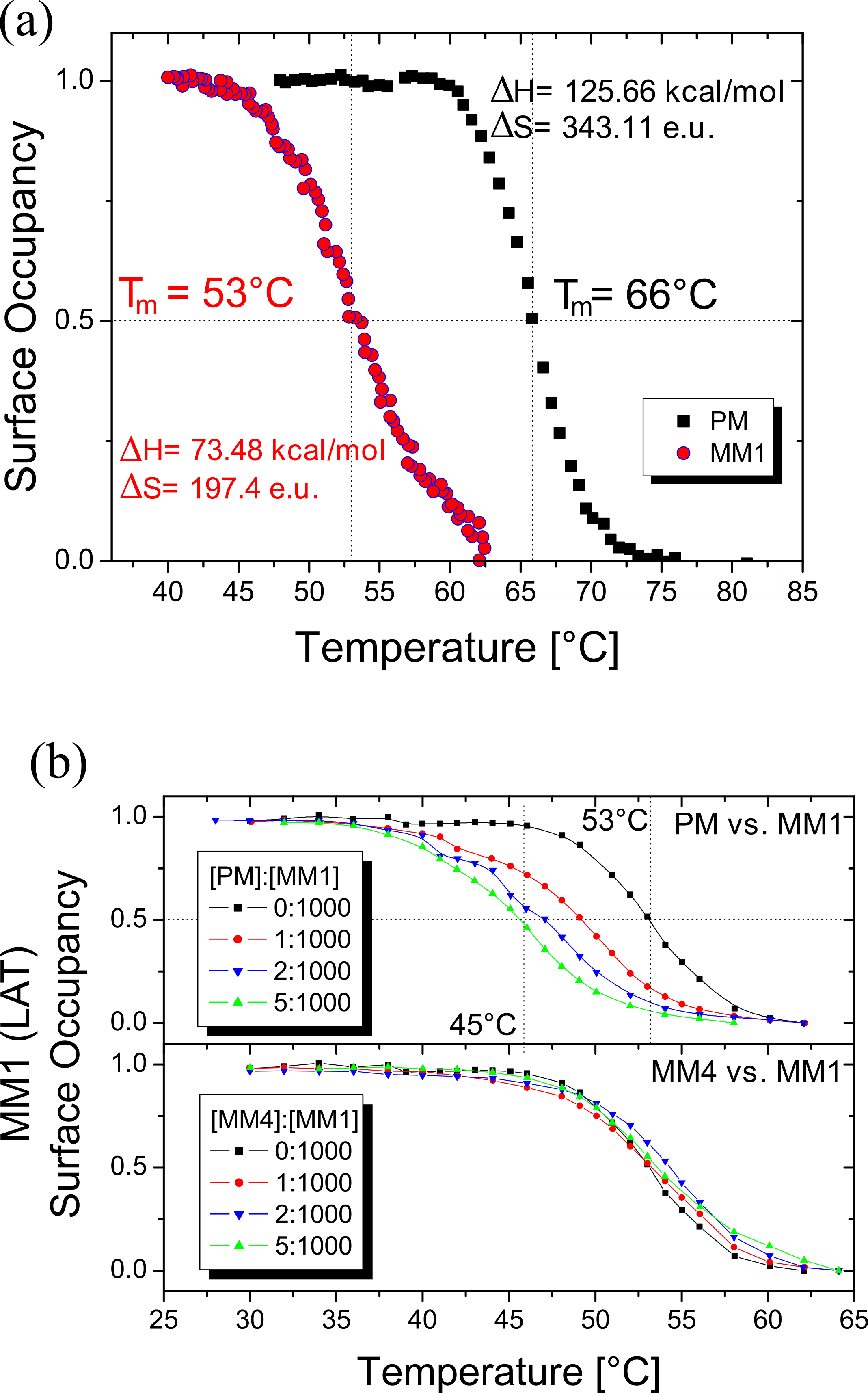}
\caption{\textbf{Melting Curves}\\ \textbf{(a)}, Fraction of target-probe complexes (the probes are surface bound) as a function of temperature for the targets PM (black)  and MM1 (red) (w/o competition) monitored using TIRF. Target concentration is 1 \ensuremath{\mu}M. From the melting curves we extract a less pronounced change in 
 $\Delta S $  and $\Delta H $  (e.u. = `entropic units' = kcal/(mol\textperiodcentered{}K)) 
upon hybridization for MM1 compared to PM (supplementary material S10 for details). Accordingly a hybridized 
MM1 possesses an increased number of entropic degrees of freedom. \textbf{(b)} 
Melting curves of MM1 in the presence of PM (upper graph, high specificity), and in 
presence of MM4 (lower graph, standard specificity) for four different competitor concentration 
ratios, each represented by a different color. In the case of `PM vs. MM1', the 
melting temperature of the MM1 decreases from 53$^{\circ}$C in the absence of the competitor 
to 45$^{\circ}$C in the presence of the PM. In the case `MM4 vs. MM1', the melting temperature remains unaffected. 
 }
 \label{fig:MeltingCurves}
\end{figure}

The melting curve, the fraction of duplexes as a function of temperature in thermodynamic 
quasi-equilibrium, yields the changes in entropy and enthalpy of the denaturation 
process (supplementary material S10 for details). The individually hybridized MM1 gains less 
enthalpy and loses less entropy in the hybridization process (Fig.~\ref{fig:MeltingCurves}(a)). In other words, probe bound, MM1 exhibits more entropic degrees of freedom than PM. Compared to PM, highly enthalpic, helicoidal binding microstates are less pronounced. 
In opposition to the standard specific pair `MM4 vs. MM1' (Fig.~\ref{fig:MeltingCurves}(b)), we observe that the presence of the high affinity target PM shifts the melting curve of the lower-affinity competitor MM1 to lower temperatures. Accordingly the presence of PM loosens MM1 binding. This increases the gap between the melting curves of the competing molecules in good agreement with the enhanced specificity. The observation exhibits the presence of a non-negligible interaction between the competitors MM1 and PM.

\section{\label{Model}Model}
\subsubsection{\label{Outline}Outline}
We previously showed that for short oligonucleotides a simple zipper with nearest neighbor binding energies constitutes an excellent model to very accurately predict double strand stability in the absence of competition \cite{Naiser:2009, Kittel:1969}. In the nearest neighbor model, only helicoidal configurations can provide nearest neighbor binding energies. Other contributions such as non specific van der Waals attractions are neglected. At the same time strand conformations such as bubbles may be introduced besides the zipper and contribute to entropy. In this framework the binding microstates of one target strand will remain unchanged if a competing, second target binds into an established target probe complex. As a result, relative statistical weights given by the respective partition functions will not change. Accordingly there is no interaction among competing strands (besides steric exclusion) and equation~\eqref{eqone} remains valid. 

In more complex situations binding of one target may alter the binding microstates of a competing target on the same probe molecule, typically through (enthalpic) deformation of the probe. If, however, the binding microstates of the competing strands are (mostly) identical, any deformation produced by one strand will change the binding microstate energies of the competitor in (almost) the same way. As a result  the ratio of the statistical weights of their bound states will not significantly change, leading again to the Boltzmann distribution (Eq.~\eqref{eqone}) a good approximation. 

To affect the validity of equation~\eqref{eqone},  the simultaneous presence of a competitor on the probe must affect binding of one target species more strongly than the other. More precisely, to reproduce our observations here, the binding microstates of targets with lower affinity (the LATs) must be affected by the presence of a competitor (either HAT or LAT) more strongly than targets with higher affinity (the HATs). 

In \cite{Mohammadi-Kambs:2019} we gave strong arguments that $\pi-\pi$ interactions \cite{Grimme:2008} have considerable statistical weight and contribute to binding conformations other than double helices. Such conformations rely on low enthalpy compared to helicoidal conformations, which means that they must exhibit important degrees of randomness. These targets must be stretched if the probe is straightened and stiffened through helix formation with another, competing strand (Fig.~\ref{fig:DNAbindingconf} c). Pulling a randomly oriented strand straight diminishes the associated entropy, creating a physical counterforce.  Accordingly, the presence of a randomly oriented competitor on the same probe will act as an entropic spring, antagonist to helicoidal probe target conformations. We find that the same energy barrier against helicoidal conformations will affect the binding constant of a mismatched strand, a LAT, much more strongly than that of a perfectly matching competitor, a HAT (this is detailed below). 

However, this mechanism by itself is not sufficient to explain our observations: to affect probe occupancy, the situation where the competitors interact must possess sufficient statistical weight. 
A perfectly matching HAT can form fully and almost fully closed helices with the probe, conformations that are not among LAT binding conformations.  The LAT does not fully occupy the probe binding site in terms of microstates. In thermodynamic equilibrium statistical weights depend on energy levels only. We understand that the HAT with its exclusive binding conformations has finite probability to occupy a probe-LAT pair. The probability that the probe is solely occupied by a LAT is reduced while three strands, HAT, LAT and probe, can meet with nonzero probability.

The topology of trifold occupation differs from pairwise binding. This entails a supplementary term in the expression of the free energy of the occupied probe, which must take into account the modified binding constants that apply to the trifold configuration. The statistical weight of the trifold interaction depends on the likelihood that both are present, that is, the product of the probabilities or concentrations of the competing target species that are attached to probes. This leads to an expression of the free energy of the probes that represents a formal analogue of the well-known Landau description of phase transitions \cite{Landau:1969}, (Eq.~\eqref{Landau}). The supplementary term describing the trifold structure will counter site occupation entropy. This yields a discontinous transition of the Gibbs free energy landscape to probe occupation by  (practically) only one of the competing molecular target species, given the condition that the equilibrium binding constant of the HAT into a LAT probe pair is of sufficient, overcritical statistical weight (Fig.~\ref{fig:Gibbslandscape}). In that case a probe bound LAT is highly likely to experience the barrier to helicoidal states (Fig.~\ref{fig:DNAbindingconf}), created by the presence of the HAT. On average the LAT-probe affinity is strongly reduced. This does not mean that the trifold configuration remains strongly populated once equilibrium is reached: it is by allowing for trifold configurations that the site entropy contribution is modified, which makes the system deviate from the Boltzmann description so that almost all probes are occupied by the better matching target, the HAT. 


\subsubsection{\label{Implementation}Model Implementation}
To picture interactions of different DNA strands in a solvent through molecular dynamics (MD-) simulations is difficult because of system size \cite{Sambriski:2009}. Accordingly, here we present a mean field approximation and show that the experimental observations can easily be reproduced quantitatively. Since we chose our sequences (Table~\ref{tab:tableone}) at random, we expect our observations to be independent from the precise oligonucleotide sequence. Because of that we limit ourselves to considering a homopolymer  that does not, however, produce overhangs upon hybridization. 

\begin{figure}[h]
\includegraphics[width=252pt, height=337pt, keepaspectratio=true]{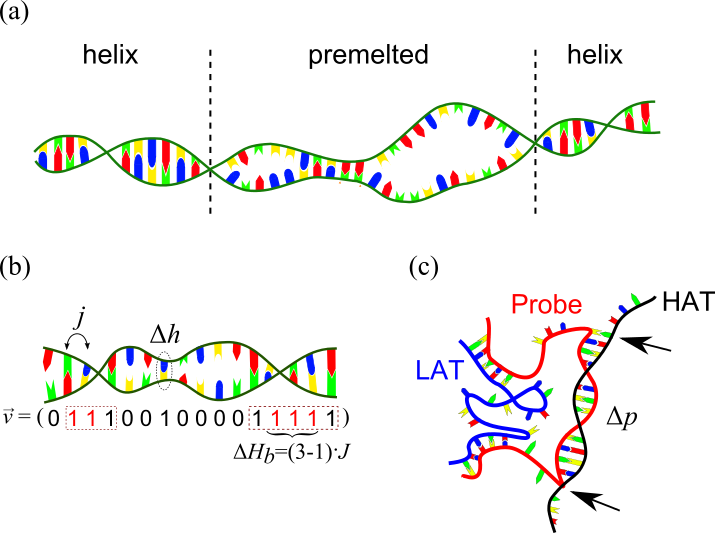}
\caption{\textbf{DNA binding configurations as considered in the model. (a)} 
Representation of intermediate and helicoidal binding configurations. 
The (premelted) bound states in the middle of the strand correspond to  the intermediate binding conformation as defined in the model. 
Base pairing is cooperative, however, the binding-enthalpy is relatively weak: this 
configuration retains many entropic degrees of freedom due to its flexibility.  
Two adjacent, closed base pairs of an intermediate configuration 
can transit to a (stiff) helicoidal configuration (left and right of the dotted lines). 
\textbf{(b)} In the model $\vec{v}  $  represents an intermediate configuration with entries 0 for each 
open base pair and 1 for each closed base pair. Each intermediate, closed base pair contributes 
by $\Delta h $. Adjacent, closed base pairs in the intermediate conformation additionally contribute to the binding enthalpy by $j$. A succession of bound bases represents a 'block' (dotted rectangles). 
Two or more adjacent pairs from a block can transit from the intermediate conformation to a helicoidal conformation (1's highlighted in red). 
The helicoidal nearest neighbor pairs contribute by $J$ to the binding enthalpy. 
This is partly offset by the stiffness of the helix, causing a loss in entropy. 
\textbf{(c)}, A trifold configuration is made of two competing target strands, 
a HAT and a LAT, bound to the same probe. If one strand, here the black HAT, transits 
to a (stiff) helicoidal conformation, this will stretch the competing strand at 
the same time. Stretching reduces the number of possible conformations, yielding a mechanical spring of entropic origin that works against helix formation. In our model we reflect the contribution of the entropic spring by a mean energy, $\Delta p$, that applies  to helicoidal conformations. Due to its non-matching base, in the presence of the barrier, compared to the HAT, the LAT is less prone to generate long helicoidal conformations that exhibit considerable statistical weight.
}
\label{fig:DNAbindingconf}
\end{figure}

In the past we provided strong evidence that for a mismatched sequence low enthalpy and high entropy configurations contribute to binding much more than double helical conformations \cite{Mohammadi-Kambs:2019}. We suggest that this leads to the observed strong deviations \cite{Altan-Bonnet:2012, Zhang:2012} from the nearest neighbor model and from related, predictive software such as Nupack. Moreover, for temperatures comparable to the melting point, the existence of so called `premelted' or `intermediate' binding conformations was shown earlier \cite{Xu:1994, Montrichok:2003, Vassili:2003, Ma:2007}. Compared to the classical double helix \cite{SantaLucia:1998}, these intermediate conformations are characterized by a lower degree of cooperativity between adjacent bases, an increased number of entropic degrees of freedom and a lower persistence length. We represent the \textit{intermediate binding configuration} by two bound single strands that maintain strand flexibility. This conformation may include open base pairs (Fig.~\ref{fig:DNAbindingconf}). We denote a particular intermediate state by a vector, $\vec{v}$, of length $N$, the number of bases of the DNA strand, with entries 0 for an open base pair and 1 for a closed one with each bond between the two strands increasing the binding enthalpy by $\Delta h$. We include a nearest-neighbor coupling energy $j$ since a base pair possesses increased closing probability if its neighbor is bound.

For each intermediate configuration $\vec{v}$ there is a set $Q(\vec{v})$  
of possible \textit{helix configurations} that the configuration can transform to. Two adjacent intermediate bonds will gain enthalpy due to stacking if they transit to such a helicoidal configuration. The double helix structure exhibits low flexibility, a reduced number of entropic degrees of freedom.
If two strands bind simultaneously and one of them passes from highly entropic, 'intermediate' configurations to establishing a helix, this requires deforming the other, competing strand. This diminishes the number of microstates of that particular strand, acting as an entropic spring against the deformation associated with helix formation (Fig.~\ref{fig:DNAbindingconf}(c)). In the following we consider the competition of a perfectly matching strand and a strand with one mismatch in its middle.

Section~\ref{FreeenergyPT}  shows how, based on the above, the corresponding partition functions can be obtained. Section~\ref{TotalFreeEnergy} sums up the different contributions to obtain the total Gibbs free energy of the system in the Landau formulation. Section~\ref{N_assessment} discusses the choice of relevant parameters. Section~\ref{Quantitative} compares the theoretical results to the experimental findings.

\subsubsection{\label{FreeenergyPT}Gibbs free energy of a probe target pair}

We write the free energy of the purely intermediate binding configuration as $\Delta G_{\vec{v} }^{n}$.
The associated helix configuration $\alpha$ generates the free energy $\Delta G_{\vec{v} ,\alpha }^{h}  $. We write 
 \begin{equation}
 \Delta G_{\vec{v} ,\alpha } =\Delta G_{\vec{v} }^{n} +\Delta G_{\vec{v} ,\alpha }^{h}.
\end{equation}
where $\alpha \in Q(\vec{v} ) $ stands for the particular helix configuration, chosen from all possible helix configurations $Q$ that can be reached through the intermediate configuration $\vec{v}$. The second term of the sum vanishes for the pure intermediate configuration, and the first vanishes otherwise.
With these definitions the value of the binding constant corresponds to the magnitude of the partition function
\begin{equation}
K^{i} =Z^{i} =\sum\limits_{\vec{v} }^{}\sum\limits_{\alpha (\vec{v} )}^{}\exp \left[ \beta \left( \Delta G_{\vec{v} ,\alpha } \right) \right]   
\label{eq:BindingConstant} 
\end{equation}
where \textit{i} designates either a HAT or a LAT target and $\beta$ the inverse of a Boltzmann factor. The first sum runs over all possible intermediate binding configurations $\vec{v}$ of the duplex. 
In the case of the HAT this is over all possible $2^{N}  $  combinations of bound 
and unbound bases in a strand of length $N $. For the LAT, however, there are 
only $2^{N-1}  $  intermediate configurations because the mismatch position remains 
unbound. In the following we consider that the LAT differs from the HAT by a single mismatch with respect to the probe. The second sum runs over all helix configurations $\alpha \in Q(\vec{v} )$.\\

\paragraph{$\Delta G_{\vec{v} }^{n} $, Gibbs free energy contribution of a (pure) intermediate binding configuration \protect\\} 

There are \textit{N}-1\textit{ }nearest-neighbor pairs in the considered DNA duplex. 
\textit{j} denotes the (constant) interaction energy between subsequent bases, 
and 
$\Delta h$
 the enthalpy gain for a bound base in the intermediate state. The enthalpy of an intermediate binding configuration, 
 $\Delta H_{\vec{v} }^{n}  $, follows an Ising description:
 \begin{equation}
\Delta H_{\vec{v} }^{n} =\Delta H_{\vec{v} }^{n} (\vec{v} )=\sum\limits_{i=1}^{N-1}j\cdot v_{i} \cdot v_{i+1} +\sum\limits_{i=1}^{N} v_{i} \cdot \Delta h 
 \end{equation}
The totally denatured configuration is the reference state, $\Delta h > 0$.

The loss in entropy upon formation of the intermediate state is estimated as follows.
An intermediate configuration has high flexibility, not all bases are bound, however, the entropy is diminished relative to
the fully denatured strands. 
 We treat the conformations of a strand as a self-avoiding walk (SAW) on a lattice \cite{Vanderzande:1998, Trapp:2011}. 
The persistence length of single strands at high salt is very short \cite{Tinland:1997, Murphy:2004, Rechendorff:2009}. As in \cite {Trapp:2011} we consider that one step on the lattice corresponds to one unbound base of the DNA strand. The number of conformations, $\Omega  $, of a SAW of length \textit{x} (for our purpose in units of numbers of bases) is
\begin{equation}
\Omega (x)=\eta ^{x} \cdot x^{\gamma -1} 
\label{eq:Omega}
\end{equation}
with $\gamma =1.157\pm 3\cdot 10^{-3}  $  the (universal) entropic exponent, while 
 $\eta =4.864 $  depends on the considered geometry \cite{Vanderzande:1998}. 
 Accordingly, the entropy change between the fully denatured configuration 
and an intermediate configuration, $\vec{v}  $, with $y $  bound bases ($=\left\| \vec{v} \right\| _{1} =\sum\limits_{i=1}^{N}v_{i}   $ ) 
is
\begin{equation}
 \Delta S_{\vec{v} }^{n} =\Delta S^{n} (y=\left\| \vec{v} \right\| _{1} )=\left\{ 
\begin{array}{l}
k_{B} N_{A} \ln\left[ \frac{\Omega (N)}{\Omega (N-y)} \right]\text{,}\text{if }y<N \\
k_{B} N_{A} \ln\left[ \Omega (N)\right]\text{, }\text{if } y=N
\end{array}
\right. 
\end{equation}
where, $k_{B}  $  is the Boltzmann constant and $N_{A}  $  is the Avogadro number. 
The intermediate configuration where all bases are bound corresponds to the largest 
entropy change. The free energy $\Delta G_{\vec{v} }^{n}  $  of a given 
intermediate configuration, $\vec{v}  $ is given by
\begin{equation}
\Delta G_{\vec{v} }^{n} =\Delta H_{\vec{v} }^{n} -T\cdot \Delta S_{\vec{v} }^{n} 
\end{equation}
where \textit{T} is the temperature.\\

\paragraph{ $\Delta G_{\vec{v} ,\alpha }^{h}$, Gibbs free energy contribution of helicoidal conformations \protect\\}
As outlined above the transition from intermediate state to a helix frees the stacking energy, $J $ for neighboring bases. For a given  set $Q(\vec{v})$ of possible 
helicoidal configurations (Fig.~\ref{fig:DNAbindingconf}(b)), for details see supplementary material S11),
if $m_{b}  $  ($2\leq m_{b} \leq l_{b}  $) bases of an individual block $b$ of $l_{b}$ successive bases in the intermediate state transit to form helices, the corresponding enthalpic gain $\Delta H_{b}  $  is
\begin{equation}
\Delta H_{b} =(m_{b} -1)\cdot J.
\label{eq:enthalpy}
\end{equation}
We obtain the enthalpy $\Delta H_{\vec{v},\alpha }^{h}$ of a given helix conformation 
 $\alpha \in Q(\vec{v} ) $by summing over all blocks $b(\alpha ) $ 
of the conformation $\vec{v}  $ 
\begin{equation}
\Delta H_{\vec{v} ,\alpha }^{h} =\sum\limits_{b(\alpha )}^{}\Delta H_{b}
\end{equation}
In case no base is in the helix conformation $\Delta H_{\vec{v} ,k}^{h} =0 $, 
and the stability of the duplex is solely governed by the intermediate binding configuration 
(see above).

The transition to a helicoidal configuration causes an entropy change, 
 $S_{\vec{v} ,\alpha }^{h}  $, due to the high persistence length of the duplex 
(the reference state being the denatured configuration). Thus the free energy 
 $\Delta G_{\vec{v} ,\alpha }^{h}  $  of a given helicoidal configuration, $\alpha $, 
is
\begin{equation}\Delta G_{\vec{v} ,\alpha }^{h} =\Delta H_{\vec{v} ,\alpha }^{h} -T\cdot S_{\vec{v} ,\alpha }^{h} \end{equation}
where
\begin{equation}S_{\vec{v} ,\alpha }^{h} =S_{\vec{v} ,\alpha }^{h} (z_{\alpha } )=k_{B} N_{A} \cdot \ln \lbrack \Omega (z_{\alpha } )]\end{equation}
is determined using Eq.~\eqref{eq:Omega}, where $z_{\alpha }$($z_{\alpha } =\sum\limits_{b(\alpha )}^{}m_{b}$) 
is the total number of bases (across all blocks) that are changed to a helix from the given 
intermediate configuration. The longest helix possesses the 
largest entropy change ($z_{\alpha } =N $) compared to the fully denatured configuration.\\

\paragraph{Gibbs free energy of a probe bound strand in the presence of a competitor \protect\\ }
Here we consider two simultaneously bound strands competing for the same probe, yielding
a trifold configuration (Fig. ~\ref{fig:DNAbindingconf}(c)). In this case we suggest that the entropic cost of 
a helicoidal conformation is increased because the occurrence of a (stiff) helix 
within such a trifold configuration requires stretching the competitor at the same 
time (Fig. ~\ref{fig:DNAbindingconf}(c)). We take this into account by imposing an energy penalty $\Delta p $  
to the helicoidal conformations in a mean field approximation, modifying  
Eq.~\eqref{eq:enthalpy}
\begin{equation}\Delta H_{b} =(m_{b} -1)\cdot J-\Delta p\end{equation}
If $(m_{b} -1)\cdot J\leq \Delta p$, 
we consider that the energy penalty prevents this conformation and neglect the contribution. The set of possible helicoidal configurations, $Q(\vec{v} ) $, depends on $\Delta p $. No energy penalty is 
applied to the fully developed helix, since the competitor 
is unable to remain bound to the same probe in that case. 


\subsubsection{\label{TotalFreeEnergy}Total Gibbs free energy of probe occupation}

Each binding site (given by a probe molecule) can be empty, occupied by a HAT only, or 
only a LAT, or simultaneously by both target species, forming a trifold. Probe-HAT-HAT 
and Probe-LAT-LAT trifolds need not be considered since these configurations also 
occur in situations without competition, and they do not alter the specificity as discussed above. 
Let $c^{HAT}  $  be the fraction of probes occupied by a HAT, and $c^{LAT}  $  
 the fraction of probes occupied by a LAT, either as single or simultaneously with 
a competing strand. The fraction of probes in the trifold state is $c^{T}  $ . 
All binding sites are occupied as a trifold if $c^{T} =1(=c^{HAT} =c^{LAT} )$. 
Considering $M $  probes in the system, the free energy per binding site has the 
form
\begin{equation}
 \frac{\Delta G}{M} =c^{HAT} \mu ^{HAT} +c^{LAT} \mu ^{LAT} -c^{T} \mu _{corr} -T\cdot \Delta S
 \label{Landau}
 \end{equation}
where $\mu ^{HAT}  $  and $\mu ^{LAT}  $  are the effective chemical potentials 
for binding of a HAT (LAT) target to the probe, $\mu _{corr}  $  
is the effective chemical potential for simultaneous binding of HAT and LAT 
to the same probe molecule and $\Delta S$ is, to a good approximation, 
the site occupation entropy.

Further $\mu ^{i} =\mu _{0}^{i} +\mu _{B}^{i}  $ with $i=\left\{ HAT,LAT\right\}  $, 
 $\mu _{0}^{i} =k_{B} N_{A} \cdot T\cdot \ln \left( K^{i} \right)  $ the standard 
Gibbs free energy that is gained if an $i $ -target hybridizes to a probe molecule 
individually and $\mu _{B}^{i} =k_{B} N_{A} \cdot T\cdot \ln \left( \lbrack i]\right)  $ the bulk concentration dependence of the chemical potential. If  $\mu_{T}^{i}  $  is the effective chemical potential for binding of an $i $ -target in a trifold configuration, and $\mu _{T}^{i} =k_{B} N_{A} \cdot T\cdot \ln (K_{T}^{i} )$  
is the standard Gibbs free energy that is gained if an $i$-target hybridizes 
to a probe to form a trifold configuration,
\begin{eqnarray}
\mu _{corr}&=&(\mu ^{HAT} -\mu ^{T,HAT} )+(\mu ^{LAT} -\mu ^{T,LAT} )\nonumber\\
 &=&(\mu _{0}^{HAT} +\mu _{B}^{HAT} -\mu _{T}^{HAT} -\mu _{B}^{HAT} )\nonumber\\
 & &+ (\mu _{0}^{LAT} +\mu _{B}^{LAT} -\mu _{T}^{LAT} -\mu _{B}^{LAT} )\nonumber\\
 &=& (\mu _{0}^{HAT} -\mu _{T}^{HAT} )+(\mu _{0}^{LAT} -\mu _{T}^{LAT} )\nonumber\\
 &=& \tilde{\mu } ^{HAT} +\tilde{\mu } ^{LAT}
 \end{eqnarray}
where, $\tilde{\mu } ^{HAT}$
 and 
$\tilde{\mu } ^{LAT}$ are the energy corrections that apply if both, HAT and LAT, are bound to one 
and the same probe.

The effective Gibbs free energy $G^{total}  $ \textsubscript{ }per site is:
\begin{eqnarray}
 \frac{G^{total} }{M}  & = & H-T\cdot S \\
 & = & c^{HAT} \cdot \mu ^{HAT} +c^{LAT} \cdot \mu ^{LAT}\nonumber\\
 && -c^{T} \cdot (\tilde{\mu } ^{HAT} +\tilde{\mu } ^{LAT} )-T\cdot S
 \label{totalGibbsfree}
 \end{eqnarray}
 The entropy $S$  is well approximated by site occupancy in units of the Boltzmann constant 
$k_{B}  $ as (the parenthesis designate binomial coefficients):
 \begin{eqnarray}
 S&=&S_{0} -k_{B} \ln \left[ 
\left(
 \begin{array}{c}
M\\
M\cdot c^{HAT} 
\end{array}
\right)
\left(
\begin{array}{c}
M \\
M\cdot c^{LAT}
\end{array}
\right)
 \right]  \\
 &\approx&S_{0}-k_{B}M\left( c^{HAT}\ln c^{HAT}+c^{LAT}\ln c^{LAT}\right)
 \end{eqnarray} 
 using the Stirling formula (as well described in related textbooks).
The constant value $S_{0}  $ is omitted in the following.

We aim at describing the problem with an interaction parameter, 
 $\kappa ^{HAT}  $  ($\kappa ^{LAT}  $ ), the conditional probability that 
a given, probe bound HAT (LAT) shares its binding site with a LAT (HAT). 
The fraction $c^{T}  $ of probes simultaneously occupied by a HAT and a LAT can be expressed  as
 $$c^{T} =c^{HAT} \cdot \kappa ^{HAT} =c^{LAT} \cdot \kappa ^{LAT} $$
 
We are interested in the case where a HAT binds to a probe already occupied by 
a LAT (corresponding to $\kappa ^{LAT}$). To simplify the notation we refer to $\kappa ^{LAT}  $  
as $\kappa  $ in the following. Eq.~\eqref{totalGibbsfree} can be expressed without explicitly referring to $ c^{T} $:
\begin{eqnarray}
\frac{G^{total}}{M}&=&c^{HAT} \cdot \mu ^{HAT} -c^{LAT} \kappa \cdot \tilde{\mu } ^{HAT}\nonumber\\
&& +c^{LAT} \cdot \mu ^{LAT} \left( 1-\frac{\kappa \cdot \tilde{\mu } ^{LAT} }{\mu ^{LAT} } \right) -T\cdot S
\end{eqnarray}
$\kappa$ depends on $ c^{HAT} $ and $ c^{LAT} $.
Since the HAT can achieve binding states that the LAT cannot reach, for a LAT occupied probe, due to the presence of free HAT in solution, there is always a non-zero probability $\kappa _{0}  $ that a HAT is part of a LAT probe pair. This implies a minimal concentration of trifolds for a concentration of bound LATs, given by
\begin{equation}
c_{min}^{T} =c^{LAT} \kappa _{0}
\end{equation}
where $\kappa _{0}  $  is a property of the competing molecules, and it depends 
on the concentration of free HATs in solution. 
$\kappa $ is first order in $c^{HAT}$. 
If all probe molecules are occupied by a HAT, all probe bound LAT molecules must share their probe with a HAT:
 $\kappa \left( c^{HAT} =1\right) =1$. The following expression of $\kappa  $, depending on $\kappa _{0}$, $c^{HAT} $  and $c^{LAT} $ fulfills these conditions.
\begin{eqnarray}
 \kappa  & = & \frac{c_{min}^{T} }{c^{LAT} } +\frac{c^{LAT} -c_{min}^{T} }{c^{LAT} (1-c_{min}^{T} )} \cdot \left( c^{HAT} -c_{min}^{T} \right)\\
 & = & \kappa _{0} +\frac{1-\kappa _{0} }{1-c^{LAT} \kappa _{0} } \cdot \left( c^{HAT} -c^{LAT} \kappa _{0} \right). 
 \end{eqnarray}
 
\subsubsection{\label{N_assessment}Numerical assessment}
We consider the HAT and the LAT as homopolymers of 16 bases 
in length. The HAT constitutes a perfect match and the LAT possesses a single, non-complementary base in its centre. The parameter values for $j $, $J $  and $\Delta h $  needed for evaluation 
of $\Delta G_{\vec{v} ,k}  $ need to be quantitatively comparable 
to the nearest-neighbor energies \cite{SantaLucia:1998}. We choose the parameters accordingly, under 
the constraint that the ratio of the theoretical binding constants $K^{HAT} /K^{LAT}  $  
corresponds to the measured value $K^{PM} /K^{MM1}  $  from single hybridization 
experiments. Table~\ref{tab:tabletwo} shows a corresponding parameter set and the resulting binding constants.
\begin{table}
\caption{\label{tab:tabletwo}Parameter set used throughout our numerical assessment, resulting binding constants at a temperature $\textit{T}  = 317 K$. The values for \textit{j}, \textit{J} and $\Delta\textit{h}$ need to be quantitatively comparable to the nearest-neighbor energies [28]. We choose the parameters in such a way that the ratio of the theoretical binding constants $ {K^{HAT} }/{K^{LAT}} $  corresponds to
 $ {K^{PM} }/{K^{MM1}} $ (as determined individually, that is, without competition) (see table ~\ref{tab:tableone} for the sequences of PM and MM1)).}
\begin{ruledtabular}
\begin{tabular}{cccccc}
 $j$ & $J$ & $\Delta h$ & $K^{HAT}$ & $K^{LAT}$ & $\frac{K^{HAT}}{K^{LAT}}$\\
$[kJ/mol]$ & $[kJ/mol]$ & $[kJ/mol]$ &&\\ \hline
$1.26$ & $7.11$ & $2.93$ & $3.20\cdot10^7$&$1.88\cdot10^6$ & $17.02$\\
 \end{tabular}
\end{ruledtabular}
\end{table}

\begin{figure}[h]
\includegraphics[width=250pt, height=382pt, keepaspectratio=true]{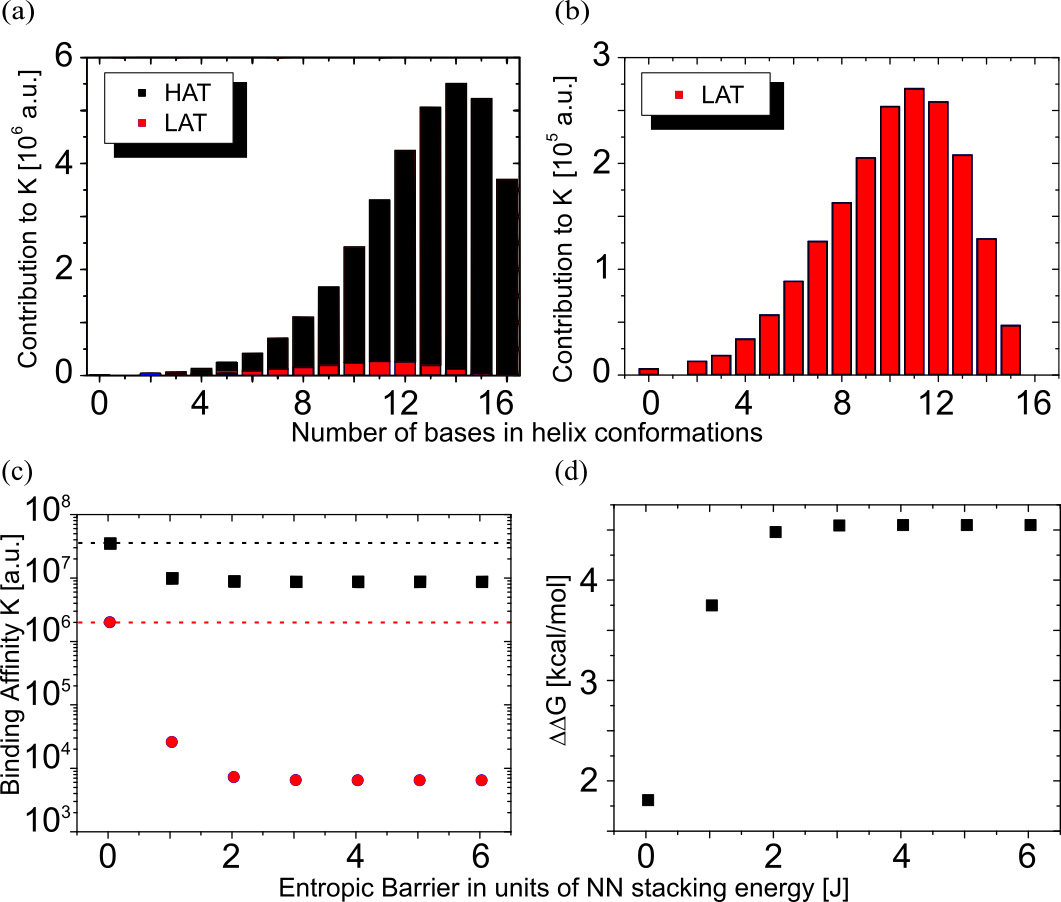}
\caption{\textbf{Numerical results based on the proposed model. (a)}, \textbf{(b)}.
Predicted contributions to the binding constant $K$ of configurations that possess a given number of base pairs in helicoidal conformation. We consider a homopolymer. \textbf{(a)}, perfect complement, 
the HAT, \textbf{(b)}, a single mismatch in the middle of the strand, 
the LAT, at 44$^{\circ}$C. Summing over all these contributions represents the binding constant $K $ (Eq.~\eqref{eq:BindingConstant}). 
Note the differences in scale: red portions in (a) represent the contributions as given in (b). The LAT generates its binding affinity through a number of short helicoidal conformations. The increased HAT binding affinity mainly stems from longer helicoidal segments (not all accessible to the LAT), and this leads to more enthalpic binding 
conformations. \textbf{(c)}, Binding affinity (Eq.~\eqref{eq:BindingConstant}) of HAT (black) and LAT (red) in a 
trifold configuration as a function of the height of the energy barrier $\Delta p $  that is 
due to the presence of a competing strand on the same probe molecule. The numerical 
assessment confirms the loss in affinity of the LAT with increasing energy barrier 
 $\Delta p $ . There is also an influence on $K^{HAT}  $, but much smaller. 
\textbf{(d)}, Change of the binding free energy difference, $\Delta \Delta G=\Delta G^{HAT} -\Delta G^{LAT}  $, 
in the presence of the competitor as a function of the energy barrier $\Delta p $.
}
\label{fig:numericalresults}
\end{figure}

(Fig. \ref{fig:numericalresults} a,b) shows the predicted contribution to the binding constant of helicoidal conformations as a function of the number of helicoidally stacked bases in pairwise binding. In the cases of only few stacked bases, the number of degenerate conformations increases roughly exponentially with the number of stacked bases. The fully helicoidal state does not make the largest contribution to the binding constant. This is due to the degeneracy of the contributions from states with slightly reduced numbers of helix conformations. This entropic contribution lifts their statistical weight above the fully closed conformation. The increased value of $K^{HAT}  $ is due to the additional helicoidal conformations that the LAT cannot produce. 
The mismatch not only prevents long helices as part of the LAT conformations, but also reduces the degeneracy of shorter helices, which eventually leads to a considerably smaller binding constant. Note also that a mismatch removes two nearest neighbor pairs (not one).
If a barrier is applied, long helices will still contribute with significant weight. The short helices of the LAT that exist to both sides of the mismatch placed in the middle of the strand, are reduced by the barrier to relatively few conformations that only weakly contribute to the partition function. Accordingly the affinity of the mismatched strand (the LAT) is considerably reduced by an energy barrier applied to helicoidal conformations as compared to the HAT. 

The numerical assessment confirms that for an effective energy barrier $\Delta p $, 
which corresponds to only very few closed base pairs in terms of free energy, the 
LAT loses several orders of magnitude of its binding affinity while the HAT affinity is affected not even by a factor of 10 (Fig.~\ref{fig:numericalresults}(c, d)). 


\subsubsection{\label{Quantitative}Comparison to experimental 
results }

\begin{figure}[h]
\includegraphics[width=250pt, height=318pt, keepaspectratio=true]{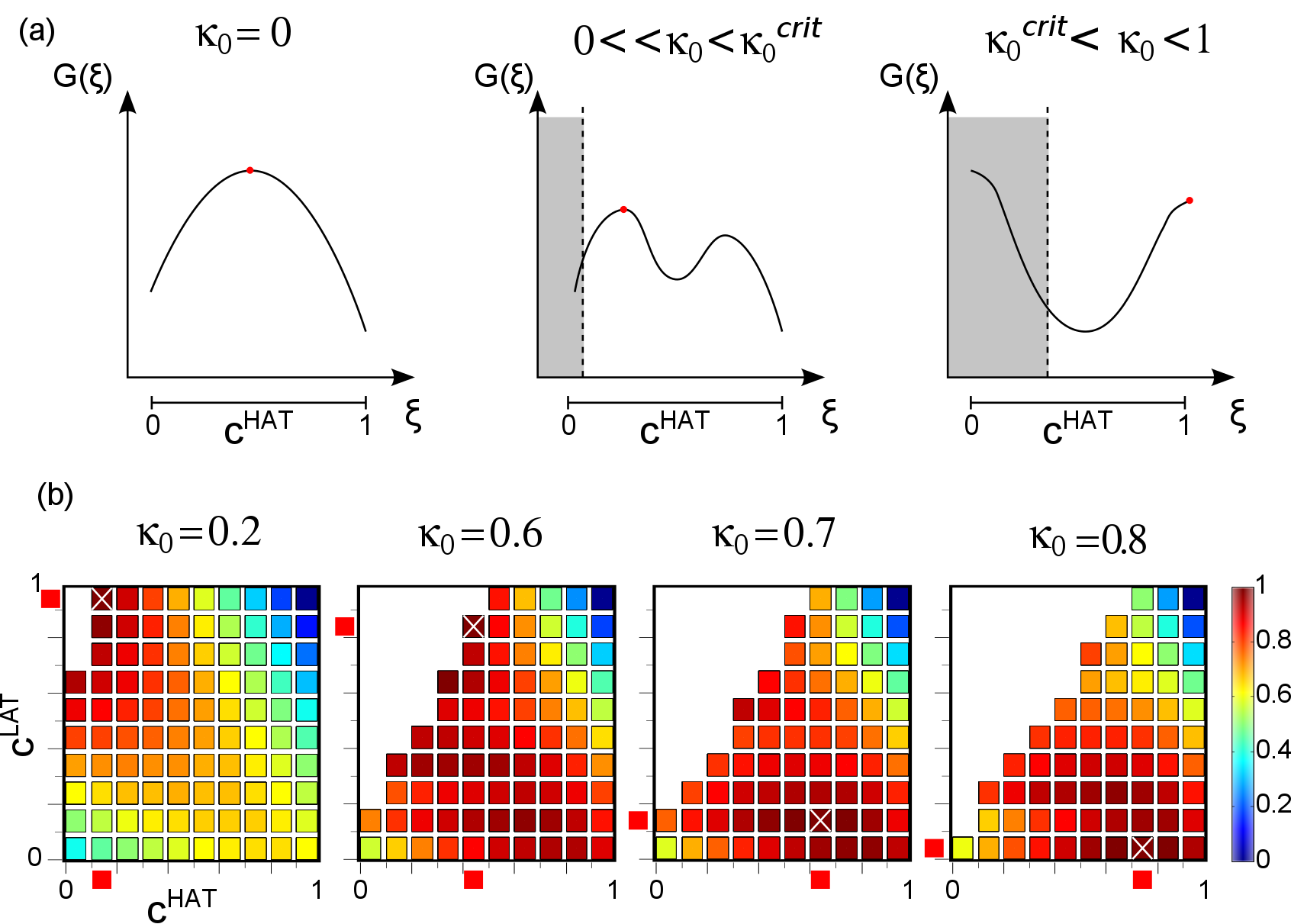}
\caption{\textbf{Gibbs free energy landscape in competitive hybridization.} \textbf{(a)}, 
Qualitative illustration of the Gibbs free energy $G(\xi)$ of the probe binding sites.$\xi$ represents $c^{HAT}$, the amount of probe bound HAT i.e. $c^{HAT}=1$ is the case where all probes are HAT occupied. The probe bound LAT concentration, $c^{LAT}$, is not shown: $c^{LAT}$ 
is such that $G(\xi)$ is maximized for every $\xi$. The overall maximum of 
$G(\xi)$ is marked with a red dot. We consider three different values of $\kappa _{0}  $, 
the conditional probability of a HAT to be part of an already formed LAT-probe pair, 
producing a trifold configuration. For $\kappa _{0} =0 $  there are no trifold 
configurations and the maximum of $G(\xi)$
is given by the Boltzmann factor, which corresponds to the concentration 
ratio${c^{HAT}}/{c^{LAT}}$. 
For $0<\kappa _{0} <\kappa _{0}^{crit}  $,  $G(\xi)$
exhibits a global maximum for slightly increased concentrations of bound LAT, $c^{LAT}$ ,
while $c^{HAT}$ is slightly reduced compared to the Boltzmann prediction. Since the number of HATs for a given number of LATs must be above $c_{min}^{T} =c^{LAT} \kappa _{0}  $, 
the shaded regions cannot be reached. For $\kappa _{0} >\kappa _{0}^{crit}  $, $G(\xi)$
switches its maximum to extreme values of $c^{HAT}$ 
and low values of $c^{LAT}$. 
\textbf{(b)}, Numerical 
result using Eq.~\eqref{totalGibbsfree}. Gibbs free energy landscape as a function of the fraction of HAT and LAT occupied probes ($c^{HAT}$ and $c^{LAT}$) for different values of $\kappa _{0}  $,  the probability for a HAT to be part of a LAT-probe pair. Here we take the experimentally determined binding constants 
of the strands PM and MM1 (table ~\ref{tab:tableone}): the HAT (PM) concentration is 5nM and the 
LAT (MM1) concentration is 1 $\mu$ M. We consider the case $\Delta p= 4J $. Maximum and minimum values of the free energy are normalized to 1 and 0. The white - X - (identified by the brown squares on x and y axis) denotes 
the maximum of the free energy landscape. Probabilities $\kappa _{0} >0.6 $  lead 
to a sharp shift in the position of the energy maximum towards the region of high 
specificity (elevated values of $c^{HAT}$ and low values of $c^{LAT}$).
}
\label{fig:Gibbslandscape}
\end{figure}

In the following, the binding constants $K^{HAT}$ (corresponding to $K^{PM}$)  and $K^{LAT}$ (corresponding to $K^{MM1}$)    
are taken according to experimental values. The experimental concentrations are 
 $\lbrack PM]_{0} =\lbrack HAT]_{0} =5\text{nM}  $  and $\lbrack MM1]_{0} =\lbrack LAT]_{0} =1\mu  $ . 
With this, for a temperature of 317 K, the energies and chemical potentials in 
Eq.  are: 
\begin{eqnarray}
\mu_0^{HAT} =&k_B N_A T \cdot \ln(K^{HAT}) =& 54.35 kJ/mol\nonumber\\
\mu_0^{LAT} =&k_B N_A T \cdot \ln(K^{LAT}) =& 47.20 kJ/mol\nonumber\\
\mu_B^{HAT} =&k_B N_A T \cdot \ln([HAT]_0) =& 50.38 kJ/mol\nonumber\\
\mu_B^{LAT} =&k_B N_A T \cdot \ln([LAT]_0) =& 36.40 kJ/mol\nonumber\\
\nonumber
\end{eqnarray}
The results of the numerical assessment (Fig.~\ref{fig:numericalresults}) suggest that for an energy barrier 
of $\Delta p=4J$ the binding affinities $K^{HAT}$
 and $K^{LAT}$ decrease by factors of 4 and 310, respectively. These reduced binding affinities 
translate to average trifold binding energies 
$\mu_T^HAT$
 and $\mu_T^LAT$
 that apply when a particular target binds to a probe molecule that is already 
occupied by the competitor:
\begin{eqnarray}
\mu_T^{HAT} =& k_B N_A T \cdot \ln(K^{HAT}/4) &= 50.67 kJ/mol\nonumber\\
\mu_T^{LAT} =& k_B N_A T \cdot \ln(K^{LAT}/310) &= 32.09 kJ/mol\nonumber\\
\nonumber
\end{eqnarray}
These energies enable us to estimate $\kappa _{0}$ to check the consistency of our approach. $\kappa _{0}$  is given by the statistical weight of the trifold configuration of the LAT occupied probe: 
\begin{equation}
\kappa _{0} =\frac{\exp(\frac{\mu ^{T,HAT}}{k_{b} N_{A} T})} {\exp(\frac{\mu ^{T,HAT}}{k_{b} N_{A} T}) +1} =53\%   
\end{equation}
with ${k_{b} N_{A}}$ the Boltzmann constant times the Avogadro number. However,  this value represents a lower bound because of our simple mean field approximation where the same energy barrier affects all the microstates in the same way (only helices with exactly one open base pair are not affected). In reality we expect helices up to a few open base pairs to prevent the competitor from constituting an entropic spring. Accordingly almost closed helices should not be affected by the barrier. These binding states, however, possess high statistical weight, a major contribution to the higher affinity of the HAT compared to the LAT (Fig. ~\ref{fig:numericalresults}(a)). As a result, $\kappa _{0}$ from above is likely to be underestimated. 
With these considerations the value of $\kappa _{0}$ above remains in good agreement with the numerical evaluation (Fig. ~\ref {fig:Gibbslandscape}) revealing that beyond a critical probability $\kappa _{0}^{crit}$ of 60-70\%, the maximum of the energy landscape sharply changes from a specificity that is well approximated by the Boltzmann factor from the average binding energies (Eq.~\eqref{eqone}) to almost exclusively HAT occupied probes.

\subsubsection{\label{Rules} Robustness of results with respect to rules and parameters}
As mentioned above, the rules governing the binding microstates of one strand in the presence of another on the same probe could be chosen differently. Allowing almost fully closed helices not to experience the entropic spring made up by the competitor certainly makes sense. 
This increases HAT affinity, however, the result will still remain below the HAT binding affinity in the absence of any penalty (Fig.~\ref{fig:numericalresults}). This means that the HAT binding constant for large energy barriers (Fig.~\ref{fig:numericalresults}) cannot be increased by more than a factor of four, even if the unreasonable assumption is made that HAT binding is not affected by the presence of the competitor at all. 

In our model the estimated free chain entropy, based on equation~\eqref{eq:Omega}, is strictly valid only for infinitely long chains, $x\rightarrow \infty  $. For finite chains it overestimates the entropic states \cite{Vanderzande:1998}. Albeit the expression still gives the correct order of magnitude and the correct trend (an increasing number of entropic freedom with larger number of unbound bases). The overestimation will lead to an overestimation of compensating enthalpic contributions, shifting  intermediate binding towards larger numbers of bound states and higher values of bond enthalpy. 
However, since the impact of the single mismatch must correspond to the experimental result, defining the average free energy of intermediate binding, the effect of a change in entropy will be very limited. In particular the asymptotic value of the binding contribution for large barriers against helicoidal conformations, that is, the contribution of pure intermediate conformations will hardly change.
In all of the above we considered that the competitors interact only through the helix conformation while the intermediate conformation remains unaffected by the presence of the competitor. Interactions at that level are unknown. Binding to the probe will be reduced due to entropic repulsion among the competing strands. A naive guess could be a reduction by a factor of two since half of the binding space is occupied by the competing strand. However, since other effects such as non-specific attraction come into play, this makes the estimation much more subtle. Here, to simplify, we consider the intermediate conformation untouched by the competitor.
From the above we understand that with our straightforward rules the resulting specificity is a conservative estimate. More subtle rules may increase the specificity, however, we expect this to change the result by less than an order of magnitude, typically a factor of two to four in specificity. Moreover, any modification of the rules to more complex ones will not fundamentally change the model, nor the qualitative behavior of the prediction. Our result is based on the contributions from microstates required to reproduce our experimental results for molecular pairs, which fixes much of the asymptotic behavior for increasing energy penalties. 

Details on the interaction between both competitors, in particular for the intermediate binding states, are unknown. In order not to become hypothetical, a detailed study is required that will have to rely on advanced experimental work reserved for later.

\section{\label{sec:level1}Discussion\protect}
In this work we determined binding constants of oligonucleotide strands, using three different techniques: FRET (in bulk), TIRF (on dendrimer surfaces) and epifluorescence (on DNA microrrays). In all studied cases the results among these techniques agreed very well. Fluorescent labels are large compared to DNA bases. 
Besides the absence of a surface this may also contribute to FRET (where three labels are required instead of one) yielding systematically lower values in the binding constants. In good agreement, in \cite{Naiser2009} the target affinity to surface bound probes was increased by about 5 kT compared to the situation in bulk.

In the first part of our work we investigated binding among molecular pairs of oligonucleotide single strands without the presence of a competitor. The binding constants as obtained for perfectly matching strands agree well with predictions from the nearest neighbor model and related software packages such as Nupack \cite{Mohammadi-Kambs:2019}. However, binding constants for strands with single mismatches differ by one to two orders of magnitude from predictions. This difference agrees well with all related experimental studies by others that we are aware of, see \cite{Zhang:2012, Bonnet:1999, Xiao:2009} and references therein. Recently we suggested that other binding states besides the double helix contribute to the binding of oligonucleotides with a mismatch. We argued that these states play only minor roles in perfectly matching strands because of the dominating statistical weight of long, helicoidal conformations \cite{Mohammadi-Kambs:2019}. Accordingly the nearest neighbor model can safely neglect such non-helicoidal binding conformations in the case of perfectly matching sequences. However, the model will fail in situations where the emergence of helicoidal conformations is reduced so that other binding conformations cannot be neglected any more.\\

Here our aim was to study target binding in competition to find out if the resulting probe occupancy could be inferred from pairwise assessments. We observed that probe occupation in competiton either obeys Boltzmann statistics (Eq.~\eqref{eqone}), based on the ratio of the binding constants from pairwise considerations, or deviates from that by orders of magnitude. In particular a two orders of magnitude overconcentrated, low affinity competitor (LAT) did not affect probe occupation in any detectable way although, following Boltzmann, the LAT is expected to cause a ten fold reduction in HAT binding affinity (Figs. \ref{fig:SBHSB}, \ref{fig:MeltingCurves}).

We do not see how in thermodynamic equilibrium this abrupt change of occupancy with only slight changes in mismatch position among the competing targets could result from anything besides a cooperative mechanism. We rule out the idea that secondary structures play a role. Along with any other possible enthalpy-based mechanism, secondary structures would lead to an amplification rather than a decrease of the observed Boltzmann deviations with the temperature of the experiment. Moreover, our strands were chosen not to possess significant secondary structures. Furtermore, our experiments are performed close to the melting temperature, 
where weak secondary structures must be negligible anyway.

For competing, independent particles that bind to an array of sites, the Boltzmann statistics of site occupancy result (to an excellent approximation) from the antagonist driving forces that emerge from site occupation entropy, which tends to mix the different strands, and site energy minimization, which favors the stronger binding competitor. Given the macromolecular nature of the competitors, here we ask by how much the site occupation entropy among the probes is modified by the fact that the LAT cannot produce all the binding microstates of the HAT, so that a LAT probe pair does not represent a 'fully occupied' binding site. As we show in section \ref{Model}, with this idea in mind we derive a free energy expression that is a formal equivalent to the Landau description of phase transitions \cite{Landau:1969}. The expression includes a product of the HAT and LAT concentrations that counteracts site occupation entropy. $\kappa _{0}  $, the probability of a HAT to be part of an existing LAT probe pair, is the critical parameter of the transition. Below $\kappa _{0}^{crit}$ our theory predicts Boltzmann statistics since site occupation entropy dominates, above $\kappa _{0}^{crit}$, the stronger binding particle prevails against the competitor, counteracting entropy. The sharpness of this transition consistently explains that in our experiments we observe either  Boltzmann statistics - or strong deviations. 

The fact that the HAT binds into the LAT probe pair with supercritical  $\kappa _{0}$ is corroborated by the reduction in melting temperature in the presence of the stronger binding competitor that is not, however, observed in the Boltzmann case (Fig.~\ref{fig:MeltingCurves}, upper panel). 

As outlined in section \ref{Model}, to bind into a LAT probe pair with supercritical probability (above $\kappa _{0}^{crit}$) the HAT must possess binding microstates of sufficient statistical weight that the LAT does not have. These binding microstates will entail a higher melting temperature of the HAT probe pair compared to the LAT probe pair. The dynamelt web server gives a good prediction of melting temperatures in situations of pairwise binding. We find that a difference of about 10 $^{\circ}$C between HAT and LAT probe pairs is observed in all cases where highly specific (i.e. non-Boltzmann) binding occurs (Fig.~\ref{fig:MeltingTemperatures}). The difference is smaller in cases of  Boltzmann occupancy: if the competing molecules are rather similar, the HAT will have relatively low tendency to bind into the LAT probe pair, and  $ \kappa _{0} $ will remain subcritical. 

A dependence on sequence of the kinetics of strand association was highlighted previously \cite{Ouldridge:2013}. The nature of Brownian fluctuations that form so called bubbles, segments where the double strand transiently separate, depends on sequence as well \cite{Altan-Bonnet:2003}. We conclude that the probability of a HAT being part of a LAT-probe pair, expressed by $\kappa _{0}  $, certainly exhibits a complex dependence on sequence. Vicinity to the critical value of $ \kappa _{0} $  naturally explains why we observe the degree of specificity to change in an abrupt and unforeseen manner with mismatch position (supplementary material S12). For our study we chose all the sequences at random, suggesting a precise sequence not to play a major role in the underlying mechanism. We therefore developed a corresponding model that relies on homopolymers.

To predict the binding affinity, besides the number of closed base pairs, the associated number of degenerate binding microstates (i.e. the entropy) must be taken into account. A strong reduction in the number of degenerate conformations is the reason why a single mismatch in the middle of the strand, although a weak perturbation in terms of enthalpy, is predicted to decrease the binding affinity by as much as three orders of magnitude for short oligonucleotides in the framework of the nearest neighbor model \cite{Mohammadi-Kambs:2019}.\\ For our theoretical description we introduced an intermediate binding state that was experimentally evidenced earlier \cite{Xu:1994, Montrichok:2003, Vassili:2003, Ma:2007, Mohammadi-Kambs:2019}. Such a state is required to reflect binding states that are not helices. The experimental binding constants for pairwise binding of mismatched strands are reproduced since the intermediate states reduce the impact of a mismatch compared to the nearest neighbor predictions. Intermediate states add entropic degrees of freedom based on weakly enthalpic interactions, independently of the presence of another strand on the  probe. 

For the reasons given in section~\ref{Model} we picture the presence of another strand on the same probe by an energy penalty that applies to helicoidal states. The relatively stronger effect that the energy penalty exerts on the LAT rather than the HAT binding constant, is again due to the reduction in the number of binding conformations. The LAT can only form short helices and their contribution to binding will be affected by the energy penalty more strongly than the long helices of the HAT. For penalties of a few base pairs already, helicoidal contributions to LAT binding become negligible with respect to the intermediate contributions, not affected by the barrier. The intermediate conformations define the binding strength that will be met asymptotically for increasing penalties. 

To test for the limit of the highly specific case experimentally, we increased the LAT concentrations as much as possible, so that the HAT-probe occupancy clearly decreased (supplementary material S13). The experiment performs by a factor of roughly two better in terms of specificity than our conservative theoretical estimation, in good agreement with the expected deviations (see model). We understand that our simple mean field model reproduces the observed quantities very accurately and consistently. 
The observed specificity exceeds by about three orders of magnitude what can be expected from pairwise, experimental assessments, exceeding at the same time the predictions from the Nupack software pack, based on the nearest neighbor model, by almost an order of magnitude. We see that in our case specificity is almost two orders of magnitude increased compared to \cite{Zhang:2012} where a competitive situation is created to increase mismatch detection. Our results equal or exceed the specificity of most optimized probes for SNP detection in competition \cite{Wang:2015} although we made no particular effort in sequence design, buffer conditions or the experimental setup, even using surface based hybridization.
Accordingly we suspect that some of the observations in \cite{Williams:2011, Wang:2015} may well be due to similar mechanisms as exposed here.

In principle nothing precludes the here described mechanism to work for much longer strands although thermodynamic equilibrium may not always be simple to reach in such a case.  We expect similar results at lower ionic concentrations though longer strands and more than one mismatch may be required to reach $\kappa _{0}^{crit}$ and observe significant deviations from the Boltzmann statistics. This is because reduced coulomb screening of the charged DNA backbone will reduce triplet formation at lower salt.

Following our interpretation, simultaneous binding of the competitors creates an 
energy barrier in analogy to conformational proofreading \cite{Savir:2007}. However, in the case 
considered here, the energy barrier is of entropic origin, created by the 
presence of a third molecule, which leads to cooperative effects. This results in almost no loss in affinity, contrary 
to conformational proofreading where a barrier of enthalpic origin is added to 
one of the two binding partners. Here the energetic cost for the observed, highly increased 
specificity is hidden in the newly created thermodynamic situation that comes with 
the presence of both competitors. The molecular behavior corresponds to logic 
`if' (the stronger binding competitor is not present) `then' (bind almost as well 
at the same spot). Thermodynamically, computing can indeed be performed at the 
simple cost of the input and output operations \cite{Schneider:1994}.

Using the here described phenomenon, very high fidelity single mismatch detection 
could be achieved in biotechnology, which could improve single nucleotide polymorphism 
(SNP) detection in medical analysis. It will be interesting to see if three-body 
mechanisms along the mechanism outlined in this paper do occur in biological systems. 
MicroRNA binding affinity to transcription RNAs has been considered a dominant 
parameter \cite{Chi:2012} for the strength of microRNA interference signaling.
Mismatches naturally occur in microRNA, and sometimes their impact is surprisingly 
strong  \cite{Mishra:2013}. Although many other molecular elements come into play in a biological cell, one 
may expect that the proposed mechanism contributes at a certain level. 

During homologous recombination, matching DNA single strands 
are combined by hybridization to repair erroneous or broken DNA strands. Following the above 
SNPs could have an increased probability of being eliminated. Note that 
hybridization in cells can be much faster than at physiological salt concentrations 
in brine \cite{Schoen:2009}. This has been attributed to coulomb screening by the highly crowded, 
charged molecular environment in the cell that relates to the buffer conditions 
used in our study.

Any macromolecule that cooperatively induces strong enthalpic changes upon binding 
to its complement may in principle follow the scheme outlined in this paper. This may even apply to proteins. The molecules will gain increased specificity in competition with competitors that bind through smaller changes in entropy (and smaller binding enthalpy) if the presence of a competitor binding to the probe with higher enthalpy affects the looser binding competitor more strongly and the probability of this interaction attains overcritical statistical weight.\\
\section{\label{Conclusion}Conclusion}
We have shown that in thermodynamic equilibrium the accuracy of macromolecular 
recognition in competition is not necessarily understood if limited to pairwise 
considerations of interactions, rather here the competing macromolecules collectively 
``proofread'' to improve their binding accuracy by orders of magnitude with almost 
no loss in affinity.

It remains to be seen if mechanisms along similar lines occur in molecular crowding 
 \cite{Ellis:2001}. Phase separations have been proposed to play a role \cite{Weber:2012} in biological 
cells. Designing ensembles of cooperative synthetic molecules, using the ideas 
discussed here, could enable the construction of self-organizing 
systems at the nano-scale with increased complexity and robustness.
\section{acknowledgements} This work was funded by the Deutsche Forschungsgemeinschaft through the collaborative research center SFB 1027.\\



\bibliography{PRX}

\begin{thebibliography}{65}%
\makeatletter
\providecommand \@ifxundefined [1]{%
 \@ifx{#1\undefined}
}%
\providecommand \@ifnum [1]{%
 \ifnum #1\expandafter \@firstoftwo
 \else \expandafter \@secondoftwo
 \fi
}%
\providecommand \@ifx [1]{%
 \ifx #1\expandafter \@firstoftwo
 \else \expandafter \@secondoftwo
 \fi
}%
\providecommand \natexlab [1]{#1}%
\providecommand \enquote  [1]{``#1''}%
\providecommand \bibnamefont  [1]{#1}%
\providecommand \bibfnamefont [1]{#1}%
\providecommand \citenamefont [1]{#1}%
\providecommand \href@noop [0]{\@secondoftwo}%
\providecommand \href [0]{\begingroup \@sanitize@url \@href}%
\providecommand \@href[1]{\@@startlink{#1}\@@href}%
\providecommand \@@href[1]{\endgroup#1\@@endlink}%
\providecommand \@sanitize@url [0]{\catcode `\\12\catcode `\$12\catcode
  `\&12\catcode `\#12\catcode `\^12\catcode `\_12\catcode `\%12\relax}%
\providecommand \@@startlink[1]{}%
\providecommand \@@endlink[0]{}%
\providecommand \url  [0]{\begingroup\@sanitize@url \@url }%
\providecommand \@url [1]{\endgroup\@href {#1}{\urlprefix }}%
\providecommand \urlprefix  [0]{URL }%
\providecommand \Eprint [0]{\href }%
\providecommand \doibase [0]{https://doi.org/}%
\providecommand \selectlanguage [0]{\@gobble}%
\providecommand \bibinfo  [0]{\@secondoftwo}%
\providecommand \bibfield  [0]{\@secondoftwo}%
\providecommand \translation [1]{[#1]}%
\providecommand \BibitemOpen [0]{}%
\providecommand \bibitemStop [0]{}%
\providecommand \bibitemNoStop [0]{.\EOS\space}%
\providecommand \EOS [0]{\spacefactor3000\relax}%
\providecommand \BibitemShut  [1]{\csname bibitem#1\endcsname}%
\let\auto@bib@innerbib\@empty
\bibitem [{\citenamefont {Lehn}(1990)}]{Lehn:1990}%
  \BibitemOpen
  \bibfield  {author} {\bibinfo {author} {\bibfnamefont {M.}~\bibnamefont
  {Lehn}},\ }\bibfield  {title} {\bibinfo {title} {Perspectives in
  supramolecular chemistry - from molecular recognition towards molecular
  information-processing and self-organization},\ }\href@noop {} {\bibfield
  {journal} {\bibinfo  {journal} {Angewandte Chemie-International Edition in
  English}\ }\textbf {\bibinfo {volume} {29}},\ \bibinfo {pages} {1304}
  (\bibinfo {year} {1990})}\BibitemShut {NoStop}%
\bibitem [{\citenamefont {Fischer}(1894)}]{Fischer:1894}%
  \BibitemOpen
  \bibfield  {author} {\bibinfo {author} {\bibfnamefont {E.}~\bibnamefont
  {Fischer}},\ }\bibfield  {title} {\bibinfo {title} {Influence of
  configuration on the action of enzymes},\ }\href@noop {} {\bibfield
  {journal} {\bibinfo  {journal} {Berichte der deutschen chemischen
  Gesellschaft}\ }\textbf {\bibinfo {volume} {27}},\ \bibinfo {pages} {2985}
  (\bibinfo {year} {1894})}\BibitemShut {NoStop}%
\bibitem [{\citenamefont {Savir}\ and\ \citenamefont
  {Tlusty}(2007)}]{Savir:2007}%
  \BibitemOpen
  \bibfield  {author} {\bibinfo {author} {\bibfnamefont {Y.}~\bibnamefont
  {Savir}}\ and\ \bibinfo {author} {\bibfnamefont {T.}~\bibnamefont {Tlusty}},\
  }\bibfield  {title} {\bibinfo {title} {Conformational proofreading: The
  impact of conformational changes on the specificity of molecular
  recognition},\ }\href@noop {} {\bibfield  {journal} {\bibinfo  {journal}
  {PLoS ONE}\ }\textbf {\bibinfo {volume} {2}},\ \bibinfo {pages} {e468}
  (\bibinfo {year} {2007})}\BibitemShut {NoStop}%
\bibitem [{\citenamefont {von Hippel}\ \emph {et~al.}(1982)\citenamefont {von
  Hippel}, \citenamefont {Kowalczykowski}, \citenamefont {Lonberg},
  \citenamefont {Newport}, \citenamefont {Paul}, \citenamefont {Stormo},\ and\
  \citenamefont {Gold}}]{Hippel:1982}%
  \BibitemOpen
  \bibfield  {author} {\bibinfo {author} {\bibfnamefont {P.}~\bibnamefont {von
  Hippel}}, \bibinfo {author} {\bibfnamefont {S.}~\bibnamefont
  {Kowalczykowski}}, \bibinfo {author} {\bibfnamefont {N.}~\bibnamefont
  {Lonberg}}, \bibinfo {author} {\bibfnamefont {J.}~\bibnamefont {Newport}},
  \bibinfo {author} {\bibfnamefont {L.}~\bibnamefont {Paul}}, \bibinfo {author}
  {\bibfnamefont {G.}~\bibnamefont {Stormo}},\ and\ \bibinfo {author}
  {\bibnamefont {Gold}},\ }\bibfield  {title} {\bibinfo {title} {Autoregulation
  of gene expression. quantitative evaluation of the expression and function of
  the bacteriophage t4 gene 32 (single-stranded dna binding) protein system},\
  }\href@noop {} {\bibfield  {journal} {\bibinfo  {journal} {Journal of
  Molecular Biology}\ }\textbf {\bibinfo {volume} {162}},\ \bibinfo {pages}
  {795} (\bibinfo {year} {1982})}\BibitemShut {NoStop}%
\bibitem [{\citenamefont {Monod}\ \emph {et~al.}(1965)\citenamefont {Monod},
  \citenamefont {Wyman},\ and\ \citenamefont {Changeux}}]{Monod:1965}%
  \BibitemOpen
  \bibfield  {author} {\bibinfo {author} {\bibfnamefont {J.}~\bibnamefont
  {Monod}}, \bibinfo {author} {\bibfnamefont {J.}~\bibnamefont {Wyman}},\ and\
  \bibinfo {author} {\bibfnamefont {J.~P.}\ \bibnamefont {Changeux}},\
  }\bibfield  {title} {\bibinfo {title} {On the nature of allosteric
  transitions: A plausible model},\ }\href@noop {} {\bibfield  {journal}
  {\bibinfo  {journal} {Journal of Molecular Biology}\ }\textbf {\bibinfo
  {volume} {12}},\ \bibinfo {pages} {88} (\bibinfo {year} {1965})}\BibitemShut
  {NoStop}%
\bibitem [{\citenamefont {Hopfield}(1974)}]{Hopfield:1974}%
  \BibitemOpen
  \bibfield  {author} {\bibinfo {author} {\bibfnamefont {J.~J.}\ \bibnamefont
  {Hopfield}},\ }\bibfield  {title} {\bibinfo {title} {Kinetic proofreading: a
  new mechanism for reducing errors in biosynthetic processes requiring high
  specificity},\ }\href@noop {} {\bibfield  {journal} {\bibinfo  {journal}
  {Proc. Natl. Acad. Sci. USA}\ }\textbf {\bibinfo {volume} {71}},\ \bibinfo
  {pages} {4135} (\bibinfo {year} {1974})}\BibitemShut {NoStop}%
\bibitem [{\citenamefont {Nin${\tilde \i}$o}(1975)}]{Ninio:1975}%
  \BibitemOpen
  \bibfield  {author} {\bibinfo {author} {\bibfnamefont {J.}~\bibnamefont
  {Nin${\tilde \i}$o}},\ }\bibfield  {title} {\bibinfo {title} {Kinetic
  amplification of enzyme discrimination},\ }\href@noop {} {\bibfield
  {journal} {\bibinfo  {journal} {Biochimie}\ }\textbf {\bibinfo {volume}
  {57}},\ \bibinfo {pages} {587} (\bibinfo {year} {1975})}\BibitemShut
  {NoStop}%
\bibitem [{\citenamefont {Bishop}\ \emph {et~al.}(2006)\citenamefont {Bishop},
  \citenamefont {Blair},\ and\ \citenamefont {Chagovetz}}]{Bishop:2006}%
  \BibitemOpen
  \bibfield  {author} {\bibinfo {author} {\bibfnamefont {J.}~\bibnamefont
  {Bishop}}, \bibinfo {author} {\bibfnamefont {S.}~\bibnamefont {Blair}},\ and\
  \bibinfo {author} {\bibfnamefont {A.~A.}\ \bibnamefont {Chagovetz}},\
  }\bibfield  {title} {\bibinfo {title} {A competitive kinetic model of nucleic
  acid surface hybridization in the presence of point mutants},\ }\href@noop {}
  {\bibfield  {journal} {\bibinfo  {journal} {Biophys. J}\ }\textbf {\bibinfo
  {volume} {90}},\ \bibinfo {pages} {831} (\bibinfo {year} {2006})}\BibitemShut
  {NoStop}%
\bibitem [{\citenamefont {Bhanot}\ \emph {et~al.}(2003)\citenamefont {Bhanot},
  \citenamefont {Louzoun}, \citenamefont {Zhu},\ and\ \citenamefont
  {DeLisi}}]{Bhanot:2003}%
  \BibitemOpen
  \bibfield  {author} {\bibinfo {author} {\bibfnamefont {G.}~\bibnamefont
  {Bhanot}}, \bibinfo {author} {\bibfnamefont {Y.}~\bibnamefont {Louzoun}},
  \bibinfo {author} {\bibfnamefont {J.}~\bibnamefont {Zhu}},\ and\ \bibinfo
  {author} {\bibfnamefont {C.}~\bibnamefont {DeLisi}},\ }\bibfield  {title}
  {\bibinfo {title} {The importance of thermodynamic equilibrium for high
  throughput gene expression arrays},\ }\href@noop {} {\bibfield  {journal}
  {\bibinfo  {journal} {Biophys. J.}\ }\textbf {\bibinfo {volume} {84}}
  (\bibinfo {year} {2003})}\BibitemShut {NoStop}%
\bibitem [{\citenamefont {Zhang}\ \emph
  {et~al.}(2012{\natexlab{a}})\citenamefont {Zhang}, \citenamefont {Chen},\
  and\ \citenamefont {Yin}}]{Zhang:2012}%
  \BibitemOpen
  \bibfield  {author} {\bibinfo {author} {\bibfnamefont {D.~Y.}\ \bibnamefont
  {Zhang}}, \bibinfo {author} {\bibfnamefont {S.~X.}\ \bibnamefont {Chen}},\
  and\ \bibinfo {author} {\bibfnamefont {P.}~\bibnamefont {Yin}},\ }\bibfield
  {title} {\bibinfo {title} {Optimizing the specificity of nucleic acid
  hybridization},\ }\href {https://doi.org/10.1038/nchem.1246} {\bibfield
  {journal} {\bibinfo  {journal} {Nature Chemistry}\ }\textbf {\bibinfo
  {volume} {4}},\ \bibinfo {pages} {208} (\bibinfo {year}
  {2012}{\natexlab{a}})}\BibitemShut {NoStop}%
\bibitem [{\citenamefont {DeVoe}\ and\ \citenamefont {Tinoco}(1962)}]{devoe62}%
  \BibitemOpen
  \bibfield  {author} {\bibinfo {author} {\bibfnamefont {H.}~\bibnamefont
  {DeVoe}}\ and\ \bibinfo {author} {\bibfnamefont {I.}~\bibnamefont {Tinoco}},\
  }\bibfield  {title} {\bibinfo {title} {The stability of helical
  polynucleotides: Base contributions},\ }\href@noop {} {\bibfield  {journal}
  {\bibinfo  {journal} {J. Mol. Biol.}\ }\textbf {\bibinfo {volume} {4}},\
  \bibinfo {pages} {500} (\bibinfo {year} {1962})}\BibitemShut {NoStop}%
\bibitem [{\citenamefont {Borer}\ \emph {et~al.}(1974)\citenamefont {Borer},
  \citenamefont {Dengler}, \citenamefont {Tinoco},\ and\ \citenamefont
  {Uhlenbeck}}]{borer74}%
  \BibitemOpen
  \bibfield  {author} {\bibinfo {author} {\bibfnamefont {P.~N.}\ \bibnamefont
  {Borer}}, \bibinfo {author} {\bibfnamefont {B.}~\bibnamefont {Dengler}},
  \bibinfo {author} {\bibfnamefont {I.}~\bibnamefont {Tinoco}},\ and\ \bibinfo
  {author} {\bibfnamefont {O.~C.}\ \bibnamefont {Uhlenbeck}},\ }\bibfield
  {title} {\bibinfo {title} {Stability of ribonucleic acid double-stranded
  helices},\ }\href@noop {} {\bibfield  {journal} {\bibinfo  {journal} {J. Mol.
  Biol.}\ }\textbf {\bibinfo {volume} {86}},\ \bibinfo {pages} {843} (\bibinfo
  {year} {1974})}\BibitemShut {NoStop}%
\bibitem [{\citenamefont {Breslauer}\ \emph {et~al.}(1986)\citenamefont
  {Breslauer}, \citenamefont {Frank}, \citenamefont {Bl{\"o}cker},\ and\
  \citenamefont {Marky}}]{breslauer86}%
  \BibitemOpen
  \bibfield  {author} {\bibinfo {author} {\bibfnamefont {K.~J.}\ \bibnamefont
  {Breslauer}}, \bibinfo {author} {\bibfnamefont {R.}~\bibnamefont {Frank}},
  \bibinfo {author} {\bibfnamefont {H.}~\bibnamefont {Bl{\"o}cker}},\ and\
  \bibinfo {author} {\bibfnamefont {L.~A.}\ \bibnamefont {Marky}},\ }\bibfield
  {title} {\bibinfo {title} {Predicting dna duplex stability from the base
  sequence},\ }\href@noop {} {\bibfield  {journal} {\bibinfo  {journal} {Proc.
  Natl. Acad. Sci. U.S.A.}\ }\textbf {\bibinfo {volume} {83}},\ \bibinfo
  {pages} {3746} (\bibinfo {year} {1986})}\BibitemShut {NoStop}%
\bibitem [{\citenamefont {SantaLucia}\ and\ \citenamefont
  {Hicks}(2004)}]{santalucia04}%
  \BibitemOpen
  \bibfield  {author} {\bibinfo {author} {\bibfnamefont {J.}~\bibnamefont
  {SantaLucia}}\ and\ \bibinfo {author} {\bibfnamefont {D.}~\bibnamefont
  {Hicks}},\ }\bibfield  {title} {\bibinfo {title} {The thermodynamics of dna
  structural motifs},\ }\href@noop {} {\bibfield  {journal} {\bibinfo
  {journal} {Annu. Rev. Biophys. Biomol. Struct.}\ }\textbf {\bibinfo {volume}
  {33}},\ \bibinfo {pages} {415} (\bibinfo {year} {2004})}\BibitemShut
  {NoStop}%
\bibitem [{\citenamefont {SantaLucia}(1998)}]{SantaLucia:1998}%
  \BibitemOpen
  \bibfield  {author} {\bibinfo {author} {\bibfnamefont {J.~A.}\ \bibnamefont
  {SantaLucia}},\ }\bibfield  {title} {\bibinfo {title} {A unified view of
  polymer, dumbbell, and oligonucleotide dna nearest-neighbor thermodynamics},\
  }\href@noop {} {\bibfield  {journal} {\bibinfo  {journal} {Proc. Natl. Acad.
  Sci. USA}\ }\textbf {\bibinfo {volume} {95}} (\bibinfo {year}
  {1998})}\BibitemShut {NoStop}%
\bibitem [{\citenamefont {Markham}\ and\ \citenamefont
  {Zuker}(2005)}]{Markham:2005}%
  \BibitemOpen
  \bibfield  {author} {\bibinfo {author} {\bibfnamefont {N.}~\bibnamefont
  {Markham}}\ and\ \bibinfo {author} {\bibfnamefont {M.}~\bibnamefont
  {Zuker}},\ }\bibfield  {title} {\bibinfo {title} {Dinamelt web server for
  nucleic acid melting prediction},\ }\href@noop {} {\bibfield  {journal}
  {\bibinfo  {journal} {Nucleic Acid Research}\ }\textbf {\bibinfo {volume}
  {33}},\ \bibinfo {pages} {W577} (\bibinfo {year} {2005})}\BibitemShut
  {NoStop}%
\bibitem [{\citenamefont {Zuker}(2003)}]{zuker03}%
  \BibitemOpen
  \bibfield  {author} {\bibinfo {author} {\bibfnamefont {M.}~\bibnamefont
  {Zuker}},\ }\bibfield  {title} {\bibinfo {title} {Mfold web server for
  nucleic acid folding and hybridization prediction},\ }\href@noop {}
  {\bibfield  {journal} {\bibinfo  {journal} {Nucleic Acids Res.}\ }\textbf
  {\bibinfo {volume} {31}},\ \bibinfo {pages} {3406} (\bibinfo {year}
  {2003})}\BibitemShut {NoStop}%
\bibitem [{\citenamefont {Zadeh}\ \emph {et~al.}(2011)\citenamefont {Zadeh},
  \citenamefont {Steenberg}, \citenamefont {Bois}, \citenamefont {Wolfe},
  \citenamefont {Pierce}, \citenamefont {Khan}, \citenamefont {Dirks},\ and\
  \citenamefont {Pierce}}]{zadeh:2011}%
  \BibitemOpen
  \bibfield  {author} {\bibinfo {author} {\bibfnamefont {J.~N.}\ \bibnamefont
  {Zadeh}}, \bibinfo {author} {\bibfnamefont {C.~D.}\ \bibnamefont
  {Steenberg}}, \bibinfo {author} {\bibfnamefont {J.~S.}\ \bibnamefont {Bois}},
  \bibinfo {author} {\bibfnamefont {B.~R.}\ \bibnamefont {Wolfe}}, \bibinfo
  {author} {\bibfnamefont {M.~B.}\ \bibnamefont {Pierce}}, \bibinfo {author}
  {\bibfnamefont {A.~R.}\ \bibnamefont {Khan}}, \bibinfo {author}
  {\bibfnamefont {R.~M.}\ \bibnamefont {Dirks}},\ and\ \bibinfo {author}
  {\bibfnamefont {N.~A.}\ \bibnamefont {Pierce}},\ }\bibfield  {title}
  {\bibinfo {title} {Nupack: Analysis and design of nucleic acid systems},\
  }\href@noop {} {\bibfield  {journal} {\bibinfo  {journal} {J. Comput. Chem.}\
  }\textbf {\bibinfo {volume} {32}},\ \bibinfo {pages} {170} (\bibinfo {year}
  {2011})}\BibitemShut {NoStop}%
\bibitem [{\citenamefont {Srinivas}\ \emph {et~al.}(2013)\citenamefont
  {Srinivas}, \citenamefont {Ouldridge}, \citenamefont {\v{S}ulc},
  \citenamefont {Schaeffer}, \citenamefont {Yurke}, \citenamefont {Louis},
  \citenamefont {Doye},\ and\ \citenamefont {Winfree}}]{srinivas13}%
  \BibitemOpen
  \bibfield  {author} {\bibinfo {author} {\bibfnamefont {N.}~\bibnamefont
  {Srinivas}}, \bibinfo {author} {\bibfnamefont {T.~E.}\ \bibnamefont
  {Ouldridge}}, \bibinfo {author} {\bibfnamefont {P.}~\bibnamefont {\v{S}ulc}},
  \bibinfo {author} {\bibfnamefont {J.~M.}\ \bibnamefont {Schaeffer}}, \bibinfo
  {author} {\bibfnamefont {B.}~\bibnamefont {Yurke}}, \bibinfo {author}
  {\bibfnamefont {A.~A.}\ \bibnamefont {Louis}}, \bibinfo {author}
  {\bibfnamefont {J.~P.~K.}\ \bibnamefont {Doye}},\ and\ \bibinfo {author}
  {\bibfnamefont {E.}~\bibnamefont {Winfree}},\ }\bibfield  {title} {\bibinfo
  {title} {On the biophysics and kinetics of toehold-mediated dna strand
  displacement},\ }\href@noop {} {\bibfield  {journal} {\bibinfo  {journal}
  {Nucleic Acids Res.}\ }\textbf {\bibinfo {volume} {41}},\ \bibinfo {pages}
  {10641} (\bibinfo {year} {2013})}\BibitemShut {NoStop}%
\bibitem [{\citenamefont {Zhang}\ and\ \citenamefont
  {Winfree}(2009)}]{zhang09}%
  \BibitemOpen
  \bibfield  {author} {\bibinfo {author} {\bibfnamefont {D.~Y.}\ \bibnamefont
  {Zhang}}\ and\ \bibinfo {author} {\bibfnamefont {E.}~\bibnamefont
  {Winfree}},\ }\bibfield  {title} {\bibinfo {title} {Control of dna strand
  displacement kinetics using toehold exchange},\ }\href@noop {} {\bibfield
  {journal} {\bibinfo  {journal} {J. Am. Chem. Soc.}\ }\textbf {\bibinfo
  {volume} {131}},\ \bibinfo {pages} {17303} (\bibinfo {year}
  {2009})}\BibitemShut {NoStop}%
\bibitem [{\citenamefont {Zhang}\ \emph
  {et~al.}(2012{\natexlab{b}})\citenamefont {Zhang}, \citenamefont {Chen},\
  and\ \citenamefont {Yin}}]{zhang12}%
  \BibitemOpen
  \bibfield  {author} {\bibinfo {author} {\bibfnamefont {D.~Y.}\ \bibnamefont
  {Zhang}}, \bibinfo {author} {\bibfnamefont {S.~X.}\ \bibnamefont {Chen}},\
  and\ \bibinfo {author} {\bibfnamefont {P.}~\bibnamefont {Yin}},\ }\bibfield
  {title} {\bibinfo {title} {Optimizing the specificity of nucleic acid
  hybridization},\ }\href@noop {} {\bibfield  {journal} {\bibinfo  {journal}
  {Nat. Chem.}\ }\textbf {\bibinfo {volume} {4}},\ \bibinfo {pages} {208}
  (\bibinfo {year} {2012}{\natexlab{b}})}\BibitemShut {NoStop}%
\bibitem [{\citenamefont {Idili}\ \emph {et~al.}(2017)\citenamefont {Idili},
  \citenamefont {Ricci},\ and\ \citenamefont
  {Vall\'{e}e-B\'{e}lisle}}]{idili17}%
  \BibitemOpen
  \bibfield  {author} {\bibinfo {author} {\bibfnamefont {A.}~\bibnamefont
  {Idili}}, \bibinfo {author} {\bibfnamefont {F.}~\bibnamefont {Ricci}},\ and\
  \bibinfo {author} {\bibfnamefont {A.}~\bibnamefont
  {Vall\'{e}e-B\'{e}lisle}},\ }\bibfield  {title} {\bibinfo {title}
  {Determining the folding and binding free energy of dna-based nanodevices and
  nanoswitches using urea titration curves},\ }\href@noop {} {\bibfield
  {journal} {\bibinfo  {journal} {Nucleic Acids Res.}\ }\textbf {\bibinfo
  {volume} {45}},\ \bibinfo {pages} {7571} (\bibinfo {year}
  {2017})}\BibitemShut {NoStop}%
\bibitem [{\citenamefont {Moreira}\ \emph {et~al.}(2005)\citenamefont
  {Moreira}, \citenamefont {You}, \citenamefont {Behlke},\ and\ \citenamefont
  {Owczarzy}}]{moreira2005}%
  \BibitemOpen
  \bibfield  {author} {\bibinfo {author} {\bibfnamefont {B.~G.}\ \bibnamefont
  {Moreira}}, \bibinfo {author} {\bibfnamefont {Y.}~\bibnamefont {You}},
  \bibinfo {author} {\bibfnamefont {M.~A.}\ \bibnamefont {Behlke}},\ and\
  \bibinfo {author} {\bibfnamefont {R.}~\bibnamefont {Owczarzy}},\ }\bibfield
  {title} {\bibinfo {title} {Effects of fluorescent dyes, quenchers, and
  dangling ends on dna duplex stability},\ }\href@noop {} {\bibfield  {journal}
  {\bibinfo  {journal} {Biochem. Biophys. Res. Commun.}\ }\textbf {\bibinfo
  {volume} {327}},\ \bibinfo {pages} {473} (\bibinfo {year}
  {2005})}\BibitemShut {NoStop}%
\bibitem [{\citenamefont {Wolk}\ \emph {et~al.}(2015)\citenamefont {Wolk},
  \citenamefont {Shoemaker}, \citenamefont {Mayfield}, \citenamefont
  {Mestdagh},\ and\ \citenamefont {Janjic}}]{wolk15}%
  \BibitemOpen
  \bibfield  {author} {\bibinfo {author} {\bibfnamefont {S.~K.}\ \bibnamefont
  {Wolk}}, \bibinfo {author} {\bibfnamefont {R.~K.}\ \bibnamefont {Shoemaker}},
  \bibinfo {author} {\bibfnamefont {W.~S.}\ \bibnamefont {Mayfield}}, \bibinfo
  {author} {\bibfnamefont {A.~L.}\ \bibnamefont {Mestdagh}},\ and\ \bibinfo
  {author} {\bibfnamefont {N.}~\bibnamefont {Janjic}},\ }\bibfield  {title}
  {\bibinfo {title} {Influence of 5-n-carboxamide modifications on the
  thermodynamic stability of oligonucleotides},\ }\href@noop {} {\bibfield
  {journal} {\bibinfo  {journal} {Nucleic Acids Res.}\ }\textbf {\bibinfo
  {volume} {43}},\ \bibinfo {pages} {9107} (\bibinfo {year}
  {2015})}\BibitemShut {NoStop}%
\bibitem [{\citenamefont {Hooyberghs}\ \emph {et~al.}(2009)\citenamefont
  {Hooyberghs}, \citenamefont {Van~Hummelen},\ and\ \citenamefont
  {Carlon}}]{Hooyberghs:2009}%
  \BibitemOpen
  \bibfield  {author} {\bibinfo {author} {\bibfnamefont {J.}~\bibnamefont
  {Hooyberghs}}, \bibinfo {author} {\bibfnamefont {P.}~\bibnamefont
  {Van~Hummelen}},\ and\ \bibinfo {author} {\bibfnamefont {E.}~\bibnamefont
  {Carlon}},\ }\bibfield  {title} {\bibinfo {title} {The effects of mismatches
  on hybridization in dna microarrays: determination of nearest neighbor
  parameters.},\ }\href@noop {} {\bibfield  {journal} {\bibinfo  {journal}
  {Nucleic Acids Res.}\ }\textbf {\bibinfo {volume} {37}},\ \bibinfo {pages}
  {e53} (\bibinfo {year} {2009})}\BibitemShut {NoStop}%
\bibitem [{\citenamefont {Huguet}\ \emph {et~al.}(2010)\citenamefont {Huguet},
  \citenamefont {Bizarro}, \citenamefont {Forns}, \citenamefont {Smith},
  \citenamefont {Bustamante},\ and\ \citenamefont {Ritort}}]{Huguet:2010}%
  \BibitemOpen
  \bibfield  {author} {\bibinfo {author} {\bibfnamefont {J.~M.}\ \bibnamefont
  {Huguet}}, \bibinfo {author} {\bibfnamefont {C.~V.}\ \bibnamefont {Bizarro}},
  \bibinfo {author} {\bibfnamefont {N.}~\bibnamefont {Forns}}, \bibinfo
  {author} {\bibfnamefont {S.~B.}\ \bibnamefont {Smith}}, \bibinfo {author}
  {\bibfnamefont {C.}~\bibnamefont {Bustamante}},\ and\ \bibinfo {author}
  {\bibfnamefont {F.}~\bibnamefont {Ritort}},\ }\bibfield  {title} {\bibinfo
  {title} {Single-molecule derivation of salt dependent base-pair free energies
  in dna},\ }\href@noop {} {\bibfield  {journal} {\bibinfo  {journal} {Proc.
  Natl. Acad. Sci. USA}\ }\textbf {\bibinfo {volume} {107}},\ \bibinfo {pages}
  {15431} (\bibinfo {year} {2010})}\BibitemShut {NoStop}%
\bibitem [{\citenamefont {Hadiwikarta}\ \emph {et~al.}(2012)\citenamefont
  {Hadiwikarta}, \citenamefont {Walter}, \citenamefont {Hooyberghs},\ and\
  \citenamefont {Carlon}}]{Hadiwikarta:2012}%
  \BibitemOpen
  \bibfield  {author} {\bibinfo {author} {\bibfnamefont {W.~W.}\ \bibnamefont
  {Hadiwikarta}}, \bibinfo {author} {\bibfnamefont {J.~C.}\ \bibnamefont
  {Walter}}, \bibinfo {author} {\bibfnamefont {J.}~\bibnamefont {Hooyberghs}},\
  and\ \bibinfo {author} {\bibfnamefont {E.}~\bibnamefont {Carlon}},\
  }\bibfield  {title} {\bibinfo {title} {Probing hybridization parameters from
  microarray experiments: nearest-neighbor model and beyond},\ }\href@noop {}
  {\bibfield  {journal} {\bibinfo  {journal} {Nucleic Acids Res.}\ }\textbf
  {\bibinfo {volume} {40}},\ \bibinfo {pages} {e138} (\bibinfo {year}
  {2012})}\BibitemShut {NoStop}%
\bibitem [{\citenamefont {Vologodskii}\ and\ \citenamefont
  {Frank-Kamenetskii}(2018)}]{Vologodskii18}%
  \BibitemOpen
  \bibfield  {author} {\bibinfo {author} {\bibfnamefont {A.}~\bibnamefont
  {Vologodskii}}\ and\ \bibinfo {author} {\bibfnamefont {M.~D.}\ \bibnamefont
  {Frank-Kamenetskii}},\ }\bibfield  {title} {\bibinfo {title} {Dna melting and
  energetics of the double helix},\ }\href
  {https://doi.org/https://doi.org/10.1016/j.plrev.2017.11.012} {\bibfield
  {journal} {\bibinfo  {journal} {Physics of Life Reviews}\ }\textbf {\bibinfo
  {volume} {25}},\ \bibinfo {pages} {1 } (\bibinfo {year} {2018})}\BibitemShut
  {NoStop}%
\bibitem [{\citenamefont {Mohammadi-Kambs}\ and\ \citenamefont
  {Ott}(2019)}]{Mohammadi-Kambs:2019}%
  \BibitemOpen
  \bibfield  {author} {\bibinfo {author} {\bibfnamefont {M.}~\bibnamefont
  {Mohammadi-Kambs}}\ and\ \bibinfo {author} {\bibfnamefont {A.}~\bibnamefont
  {Ott}},\ }\bibfield  {title} {\bibinfo {title} {Dna oligomer binding in
  competition exhibits cooperativity},\ }\href@noop {} {\bibfield  {journal}
  {\bibinfo  {journal} {New J. Phys.}\ }\textbf {\bibinfo {volume} {21}},\
  \bibinfo {pages} {113027} (\bibinfo {year} {2019})}\BibitemShut {NoStop}%
\bibitem [{\citenamefont {Lee}\ \emph {et~al.}(2014)\citenamefont {Lee},
  \citenamefont {D.}, \citenamefont {Nieman}, \citenamefont {C.},\ and\
  \citenamefont {Lanzaro}}]{Lee:2014}%
  \BibitemOpen
  \bibfield  {author} {\bibinfo {author} {\bibfnamefont {Y.}~\bibnamefont
  {Lee}}, \bibinfo {author} {\bibfnamefont {M.~C.}\ \bibnamefont {D.}},
  \bibinfo {author} {\bibnamefont {Nieman}}, \bibinfo {author} {\bibnamefont
  {C.}},\ and\ \bibinfo {author} {\bibfnamefont {G.}~\bibnamefont {Lanzaro}},\
  }\bibfield  {title} {\bibinfo {title} {A new multiplex snp genotyping assay
  for detecting hybridization and introgression between the m and s molecular
  forms of anopheles gabiae},\ }\href@noop {} {\bibfield  {journal} {\bibinfo
  {journal} {Mol. Ecol. Res.}\ }\textbf {\bibinfo {volume} {14}},\ \bibinfo
  {pages} {297} (\bibinfo {year} {2014})}\BibitemShut {NoStop}%
\bibitem [{\citenamefont {Bonnet}\ \emph {et~al.}(1999)\citenamefont {Bonnet},
  \citenamefont {Tyagi}, \citenamefont {Libchaber},\ and\ \citenamefont
  {Kramer}}]{Bonnet:1999}%
  \BibitemOpen
  \bibfield  {author} {\bibinfo {author} {\bibfnamefont {G.}~\bibnamefont
  {Bonnet}}, \bibinfo {author} {\bibfnamefont {S.}~\bibnamefont {Tyagi}},
  \bibinfo {author} {\bibfnamefont {A.}~\bibnamefont {Libchaber}},\ and\
  \bibinfo {author} {\bibfnamefont {F.}~\bibnamefont {Kramer}},\ }\bibfield
  {title} {\bibinfo {title} {Thermodynamic basis of the enhanced specificity of
  structured dna probes},\ }\href@noop {} {\bibfield  {journal} {\bibinfo
  {journal} {Proc. Natl. Acad. Sci. U.S.A.}\ }\textbf {\bibinfo {volume}
  {96}},\ \bibinfo {pages} {6171} (\bibinfo {year} {1999})}\BibitemShut
  {NoStop}%
\bibitem [{\citenamefont {Altan-Bonnet}\ and\ \citenamefont
  {Kramer}(2012)}]{Altan-Bonnet:2012}%
  \BibitemOpen
  \bibfield  {author} {\bibinfo {author} {\bibfnamefont {G.}~\bibnamefont
  {Altan-Bonnet}}\ and\ \bibinfo {author} {\bibfnamefont {F.~R.}\ \bibnamefont
  {Kramer}},\ }\bibfield  {title} {\bibinfo {title} {Robust sequence
  discrimination},\ }\href@noop {} {\bibfield  {journal} {\bibinfo  {journal}
  {Nature Chem.}\ }\textbf {\bibinfo {volume} {4}} (\bibinfo {year}
  {2012})}\BibitemShut {NoStop}%
\bibitem [{\citenamefont {Wang}\ and\ \citenamefont {Zhang}(2015)}]{Wang:2015}%
  \BibitemOpen
  \bibfield  {author} {\bibinfo {author} {\bibfnamefont {J.~S.}\ \bibnamefont
  {Wang}}\ and\ \bibinfo {author} {\bibfnamefont {D.~Y.}\ \bibnamefont
  {Zhang}},\ }\bibfield  {title} {\bibinfo {title} {Simulation-guided dna probe
  design for consistently ultraspecific hybridization},\ }\href
  {https://doi.org/10.1038/nchem.2266} {\bibfield  {journal} {\bibinfo
  {journal} {Nature Chemistry}\ }\textbf {\bibinfo {volume} {7}},\ \bibinfo
  {pages} {545} (\bibinfo {year} {2015})}\BibitemShut {NoStop}%
\bibitem [{\citenamefont {Brecevic}\ \emph {et~al.}(2006)\citenamefont
  {Brecevic}, \citenamefont {Michel}, \citenamefont {Starke}, \citenamefont
  {M{\"u}ller}, \citenamefont {Kosyakova}, \citenamefont {Mrasek},
  \citenamefont {Weise},\ and\ \citenamefont {Liehr}}]{Brecevic:2006}%
  \BibitemOpen
  \bibfield  {author} {\bibinfo {author} {\bibfnamefont {L.}~\bibnamefont
  {Brecevic}}, \bibinfo {author} {\bibfnamefont {S.}~\bibnamefont {Michel}},
  \bibinfo {author} {\bibfnamefont {H.}~\bibnamefont {Starke}}, \bibinfo
  {author} {\bibfnamefont {K.}~\bibnamefont {M{\"u}ller}}, \bibinfo {author}
  {\bibfnamefont {N.}~\bibnamefont {Kosyakova}}, \bibinfo {author}
  {\bibfnamefont {K.}~\bibnamefont {Mrasek}}, \bibinfo {author} {\bibfnamefont
  {A.}~\bibnamefont {Weise}},\ and\ \bibinfo {author} {\bibfnamefont
  {T.}~\bibnamefont {Liehr}},\ }\bibfield  {title} {\bibinfo {title}
  {Multicolor fish used for the characterization of small supernumerary marker
  chromosomes (ssmc) in commercially available immortalized cell lines},\
  }\href@noop {} {\bibfield  {journal} {\bibinfo  {journal} {Cytogenetic and
  Genome Research}\ }\textbf {\bibinfo {volume} {114}},\ \bibinfo {pages} {319}
  (\bibinfo {year} {2006})}\BibitemShut {NoStop}%
\bibitem [{\citenamefont {Zhang}\ \emph {et~al.}(2005)\citenamefont {Zhang},
  \citenamefont {Hammer},\ and\ \citenamefont {Graves}}]{Zhang:2005}%
  \BibitemOpen
  \bibfield  {author} {\bibinfo {author} {\bibfnamefont {D.~Y.}\ \bibnamefont
  {Zhang}}, \bibinfo {author} {\bibfnamefont {D.~A.}\ \bibnamefont {Hammer}},\
  and\ \bibinfo {author} {\bibfnamefont {D.~J.}\ \bibnamefont {Graves}},\
  }\bibfield  {title} {\bibinfo {title} {Competitive hybridization kinetics
  reveals unexpected behavior patterns},\ }\href@noop {} {\bibfield  {journal}
  {\bibinfo  {journal} {Biophys. J.}\ }\textbf {\bibinfo {volume} {89}},\
  \bibinfo {pages} {2950} (\bibinfo {year} {2005})}\BibitemShut {NoStop}%
\bibitem [{\citenamefont {Horne}\ \emph {et~al.}(2006)\citenamefont {Horne},
  \citenamefont {Fish},\ and\ \citenamefont {Benight}}]{Horne:2006}%
  \BibitemOpen
  \bibfield  {author} {\bibinfo {author} {\bibfnamefont {M.}~\bibnamefont
  {Horne}}, \bibinfo {author} {\bibfnamefont {D.}~\bibnamefont {Fish}},\ and\
  \bibinfo {author} {\bibfnamefont {A.}~\bibnamefont {Benight}},\ }\bibfield
  {title} {\bibinfo {title} {Statistical thermodynamics and kinetics of dna
  multiplex hybridization reactions},\ }\href@noop {} {\bibfield  {journal}
  {\bibinfo  {journal} {Biophys. J.}\ }\textbf {\bibinfo {volume} {91}},\
  \bibinfo {pages} {4133} (\bibinfo {year} {2006})}\BibitemShut {NoStop}%
\bibitem [{\citenamefont {Bishop}\ \emph {et~al.}(2008)\citenamefont {Bishop},
  \citenamefont {Chagovets},\ and\ \citenamefont {Blair}}]{Bishop:2008}%
  \BibitemOpen
  \bibfield  {author} {\bibinfo {author} {\bibfnamefont {J.}~\bibnamefont
  {Bishop}}, \bibinfo {author} {\bibfnamefont {A.~M.}\ \bibnamefont
  {Chagovets}},\ and\ \bibinfo {author} {\bibfnamefont {S.}~\bibnamefont
  {Blair}},\ }\bibfield  {title} {\bibinfo {title} {Kinetics of multiplex
  hybridization: Mechanism and implications},\ }\href@noop {} {\bibfield
  {journal} {\bibinfo  {journal} {Biophys. J.}\ }\textbf {\bibinfo {volume}
  {94}},\ \bibinfo {pages} {1726} (\bibinfo {year} {2008})}\BibitemShut
  {NoStop}%
\bibitem [{\citenamefont {Cherepinsky}\ \emph {et~al.}(2010)\citenamefont
  {Cherepinsky}, \citenamefont {Hashmi},\ and\ \citenamefont
  {Mishra}}]{Cherepinsky:2010}%
  \BibitemOpen
  \bibfield  {author} {\bibinfo {author} {\bibfnamefont {V.}~\bibnamefont
  {Cherepinsky}}, \bibinfo {author} {\bibfnamefont {G.}~\bibnamefont
  {Hashmi}},\ and\ \bibinfo {author} {\bibfnamefont {B.}~\bibnamefont
  {Mishra}},\ }\bibfield  {title} {\bibinfo {title} {Competitive hybridization
  models},\ }\href@noop {} {\bibfield  {journal} {\bibinfo  {journal} {Phys.
  Rev. E}\ }\textbf {\bibinfo {volume} {82}},\ \bibinfo {pages} {051914}
  (\bibinfo {year} {2010})}\BibitemShut {NoStop}%
\bibitem [{\citenamefont {Williams}\ \emph {et~al.}(2011)\citenamefont
  {Williams}, \citenamefont {Blair}, \citenamefont {Chagovetz},\ and\
  \citenamefont {Benight}}]{Williams:2011}%
  \BibitemOpen
  \bibfield  {author} {\bibinfo {author} {\bibfnamefont {L.}~\bibnamefont
  {Williams}}, \bibinfo {author} {\bibfnamefont {S.}~\bibnamefont {Blair}},
  \bibinfo {author} {\bibfnamefont {D.}~\bibnamefont {Chagovetz}, \bibfnamefont
  {A.and~Fish}},\ and\ \bibinfo {author} {\bibfnamefont {A.}~\bibnamefont
  {Benight}},\ }\bibfield  {title} {\bibinfo {title} {The paradox of multiple
  dna melting on a surface},\ }\href@noop {} {\bibfield  {journal} {\bibinfo
  {journal} {Analytical Biochemistry}\ }\textbf {\bibinfo {volume} {409}},\
  \bibinfo {pages} {150} (\bibinfo {year} {2011})}\BibitemShut {NoStop}%
\bibitem [{\citenamefont {Naiser}\ \emph {et~al.}(2009)\citenamefont {Naiser},
  \citenamefont {Kayser}, \citenamefont {Mai}, \citenamefont {Michel},\ and\
  \citenamefont {Ott}}]{Naiser:2009}%
  \BibitemOpen
  \bibfield  {author} {\bibinfo {author} {\bibfnamefont {T.}~\bibnamefont
  {Naiser}}, \bibinfo {author} {\bibfnamefont {J.}~\bibnamefont {Kayser}},
  \bibinfo {author} {\bibfnamefont {T.}~\bibnamefont {Mai}}, \bibinfo {author}
  {\bibfnamefont {W.}~\bibnamefont {Michel}},\ and\ \bibinfo {author}
  {\bibfnamefont {A.}~\bibnamefont {Ott}},\ }\bibfield  {title} {\bibinfo
  {title} {Stability of a surface-bound oligonucleotide duplex inferred from
  molecular dynamics: a study of single nucleotide defects using dna
  microarrays},\ }\href@noop {} {\bibfield  {journal} {\bibinfo  {journal}
  {Phys. Rev. Lett.}\ }\textbf {\bibinfo {volume} {102}},\ \bibinfo {pages}
  {218301} (\bibinfo {year} {2009})}\BibitemShut {NoStop}%
\bibitem [{\citenamefont {Tr{\'e}visol}\ \emph {et~al.}(2003)\citenamefont
  {Tr{\'e}visol}, \citenamefont {Le-Berre},\ and\ \citenamefont
  {Leclaire}}]{Trevisol:2003}%
  \BibitemOpen
  \bibfield  {author} {\bibinfo {author} {\bibfnamefont {E.}~\bibnamefont
  {Tr{\'e}visol}}, \bibinfo {author} {\bibfnamefont {A.}~\bibnamefont
  {Le-Berre}},\ and\ \bibinfo {author} {\bibfnamefont {J.}~\bibnamefont
  {Leclaire}},\ }\bibfield  {title} {\bibinfo {title} {Dendrislides,
  dendrichips: a simple chemical functionalization of glass slides with
  phosphorous dendrimers as an effective means for the preparation of
  biochips},\ }\href@noop {} {\bibfield  {journal} {\bibinfo  {journal} {New J.
  Chem}\ }\textbf {\bibinfo {volume} {27}},\ \bibinfo {pages} {1713} (\bibinfo
  {year} {2003})}\BibitemShut {NoStop}%
\bibitem [{\citenamefont {Naiser}(2008)}]{Naiser:2008}%
  \BibitemOpen
  \bibfield  {author} {\bibinfo {author} {\bibfnamefont {T.}~\bibnamefont
  {Naiser}},\ }\emph {\bibinfo {title} {Characterization of Oligonucleotide
  Microarray Hybridization Microarray Fabrication by Light-Directed in situ
  Synthesis}},\ \href@noop {} {Ph.D. thesis},\ \bibinfo  {school}
  {Universit{\"a}t Bayreuth} (\bibinfo {year} {2008})\BibitemShut {NoStop}%
\bibitem [{\citenamefont {Michel}\ \emph {et~al.}(2007)\citenamefont {Michel},
  \citenamefont {Mai}, \citenamefont {Naiser},\ and\ \citenamefont
  {Ott}}]{Michel:2007}%
  \BibitemOpen
  \bibfield  {author} {\bibinfo {author} {\bibfnamefont {W.}~\bibnamefont
  {Michel}}, \bibinfo {author} {\bibfnamefont {T.}~\bibnamefont {Mai}},
  \bibinfo {author} {\bibfnamefont {T.}~\bibnamefont {Naiser}},\ and\ \bibinfo
  {author} {\bibfnamefont {A.}~\bibnamefont {Ott}},\ }\bibfield  {title}
  {\bibinfo {title} {Optical study of dna surface hybridization reveals dna
  surface density as a key parameter for microarray hybridization kinetics},\
  }\href@noop {} {\bibfield  {journal} {\bibinfo  {journal} {Biophysical
  Journal}\ }\textbf {\bibinfo {volume} {92}},\ \bibinfo {pages} {999}
  (\bibinfo {year} {2007})}\BibitemShut {NoStop}%
\bibitem [{\citenamefont {Kittel}(1969)}]{Kittel:1969}%
  \BibitemOpen
  \bibfield  {author} {\bibinfo {author} {\bibfnamefont {C.}~\bibnamefont
  {Kittel}},\ }\bibfield  {title} {\bibinfo {title} {Phase transition of a
  molecular zipper},\ }\href@noop {} {\bibfield  {journal} {\bibinfo  {journal}
  {Am. J. Phys.}\ }\textbf {\bibinfo {volume} {37}},\ \bibinfo {pages} {917}
  (\bibinfo {year} {1969})}\BibitemShut {NoStop}%
\bibitem [{\citenamefont {Grimme}(2008)}]{Grimme:2008}%
  \BibitemOpen
  \bibfield  {author} {\bibinfo {author} {\bibfnamefont {S.}~\bibnamefont
  {Grimme}},\ }\bibfield  {title} {\bibinfo {title} {Do special noncovalent
  $\pi$ - $\pi$ stacking interactions really exist?},\ }\href@noop {}
  {\bibfield  {journal} {\bibinfo  {journal} {Angew. Chem. Int. Ed.}\ }\textbf
  {\bibinfo {volume} {47}},\ \bibinfo {pages} {3430} (\bibinfo {year}
  {2008})}\BibitemShut {NoStop}%
\bibitem [{\citenamefont {Landau}(1969)}]{Landau:1969}%
  \BibitemOpen
  \bibfield  {author} {\bibinfo {author} {\bibfnamefont {L.}~\bibnamefont
  {Landau}},\ }\href@noop {} {\emph {\bibinfo {title} {Collected Papers}}}\
  (\bibinfo  {publisher} {Nauka, Moscow},\ \bibinfo {year} {1969})\BibitemShut
  {NoStop}%
\bibitem [{\citenamefont {Sambriski}\ \emph {et~al.}(2009)\citenamefont
  {Sambriski}, \citenamefont {Schwartz},\ and\ \citenamefont
  {de~Pablo}}]{Sambriski:2009}%
  \BibitemOpen
  \bibfield  {author} {\bibinfo {author} {\bibfnamefont {E.~J.}\ \bibnamefont
  {Sambriski}}, \bibinfo {author} {\bibfnamefont {D.~C.}\ \bibnamefont
  {Schwartz}},\ and\ \bibinfo {author} {\bibfnamefont {J.~J.}\ \bibnamefont
  {de~Pablo}},\ }\bibfield  {title} {\bibinfo {title} {Uncovering pathways in
  dna oligonucleotide hybridization via transition state analysis},\
  }\href@noop {} {\bibfield  {journal} {\bibinfo  {journal} {Proc. Natl. Acad.
  Sci. USA}\ }\textbf {\bibinfo {volume} {106}},\ \bibinfo {pages} {18125}
  (\bibinfo {year} {2009})}\BibitemShut {NoStop}%
\bibitem [{\citenamefont {Xu}\ \emph {et~al.}(1994)\citenamefont {Xu},
  \citenamefont {Evans},\ and\ \citenamefont {Nordlund}}]{Xu:1994}%
  \BibitemOpen
  \bibfield  {author} {\bibinfo {author} {\bibfnamefont {D.}~\bibnamefont
  {Xu}}, \bibinfo {author} {\bibfnamefont {K.~O.}\ \bibnamefont {Evans}},\ and\
  \bibinfo {author} {\bibfnamefont {T.~M.}\ \bibnamefont {Nordlund}},\
  }\bibfield  {title} {\bibinfo {title} {Melting and premelting transitions of
  an oligomer measured by dna base fluorescence and absorption},\ }\href@noop
  {} {\bibfield  {journal} {\bibinfo  {journal} {Biochemistry}\ }\textbf
  {\bibinfo {volume} {33}},\ \bibinfo {pages} {9592} (\bibinfo {year}
  {1994})}\BibitemShut {NoStop}%
\bibitem [{\citenamefont {Montrichok}\ \emph {et~al.}(2003)\citenamefont
  {Montrichok}, \citenamefont {Gruner},\ and\ \citenamefont
  {Zocchi}}]{Montrichok:2003}%
  \BibitemOpen
  \bibfield  {author} {\bibinfo {author} {\bibfnamefont {A.}~\bibnamefont
  {Montrichok}}, \bibinfo {author} {\bibfnamefont {G.}~\bibnamefont {Gruner}},\
  and\ \bibinfo {author} {\bibfnamefont {G.}~\bibnamefont {Zocchi}},\
  }\bibfield  {title} {\bibinfo {title} {Trapping intermediates in the melting
  transition of {DNA} oligomers},\ }\href
  {https://doi.org/10.1209/epl/i2003-00417-3} {\bibfield  {journal} {\bibinfo
  {journal} {Europhysics Letters ({EPL})}\ }\textbf {\bibinfo {volume} {62}},\
  \bibinfo {pages} {452} (\bibinfo {year} {2003})}\BibitemShut {NoStop}%
\bibitem [{\citenamefont {Ivanov}\ \emph {et~al.}(2003)\citenamefont {Ivanov},
  \citenamefont {Grzeskowiak},\ and\ \citenamefont {Zocchi}}]{Vassili:2003}%
  \BibitemOpen
  \bibfield  {author} {\bibinfo {author} {\bibfnamefont {V.}~\bibnamefont
  {Ivanov}}, \bibinfo {author} {\bibfnamefont {K.}~\bibnamefont
  {Grzeskowiak}},\ and\ \bibinfo {author} {\bibfnamefont {G.}~\bibnamefont
  {Zocchi}},\ }\bibfield  {title} {\bibinfo {title} {Evidence for an
  intermediate state in the b-to-z transition of dna},\ }\href
  {https://doi.org/10.1021/jp035593p} {\bibfield  {journal} {\bibinfo
  {journal} {The Journal of Physical Chemistry B}\ }\textbf {\bibinfo {volume}
  {107}},\ \bibinfo {pages} {12847} (\bibinfo {year} {2003})}\BibitemShut
  {NoStop}%
\bibitem [{\citenamefont {Ma}\ \emph {et~al.}(2007)\citenamefont {Ma},
  \citenamefont {Wan}, \citenamefont {Wu},\ and\ \citenamefont
  {Zewail}}]{Ma:2007}%
  \BibitemOpen
  \bibfield  {author} {\bibinfo {author} {\bibfnamefont {H.}~\bibnamefont
  {Ma}}, \bibinfo {author} {\bibfnamefont {C.}~\bibnamefont {Wan}}, \bibinfo
  {author} {\bibfnamefont {A.}~\bibnamefont {Wu}},\ and\ \bibinfo {author}
  {\bibfnamefont {A.~H.}\ \bibnamefont {Zewail}},\ }\bibfield  {title}
  {\bibinfo {title} {Dna folding and melting observed in real time redefine the
  energy landscape},\ }\href {https://doi.org/10.1073/pnas.0610028104}
  {\bibfield  {journal} {\bibinfo  {journal} {Proceedings of the National
  Academy of Sciences}\ }\textbf {\bibinfo {volume} {104}},\ \bibinfo {pages}
  {712} (\bibinfo {year} {2007})}\BibitemShut {NoStop}%
\bibitem [{\citenamefont {Vanderzande}(1998)}]{Vanderzande:1998}%
  \BibitemOpen
  \bibfield  {author} {\bibinfo {author} {\bibfnamefont {C.}~\bibnamefont
  {Vanderzande}},\ }\href@noop {} {\emph {\bibinfo {title} {Lattice models of
  polymers}}}\ (\bibinfo  {publisher} {Cambridge University Press},\ \bibinfo
  {year} {1998})\BibitemShut {NoStop}%
\bibitem [{\citenamefont {Trapp}\ \emph {et~al.}(2011)\citenamefont {Trapp},
  \citenamefont {Schenkelberger},\ and\ \citenamefont {Ott}}]{Trapp:2011}%
  \BibitemOpen
  \bibfield  {author} {\bibinfo {author} {\bibfnamefont {C.}~\bibnamefont
  {Trapp}}, \bibinfo {author} {\bibfnamefont {M.}~\bibnamefont
  {Schenkelberger}},\ and\ \bibinfo {author} {\bibfnamefont {A.}~\bibnamefont
  {Ott}},\ }\bibfield  {title} {\bibinfo {title} {Stability of double-stranded
  oligonucleotide dna with a bulged loop: a microarray study},\ }\href@noop {}
  {\bibfield  {journal} {\bibinfo  {journal} {BMC Biophysics}\ }\textbf
  {\bibinfo {volume} {4}} (\bibinfo {year} {2011})}\BibitemShut {NoStop}%
\bibitem [{\citenamefont {Tinland}\ \emph {et~al.}(1997)\citenamefont
  {Tinland}, \citenamefont {Pluen}, \citenamefont {Sturm},\ and\ \citenamefont
  {Weill}}]{Tinland:1997}%
  \BibitemOpen
  \bibfield  {author} {\bibinfo {author} {\bibfnamefont {B.}~\bibnamefont
  {Tinland}}, \bibinfo {author} {\bibfnamefont {A.}~\bibnamefont {Pluen}},
  \bibinfo {author} {\bibfnamefont {J.}~\bibnamefont {Sturm}},\ and\ \bibinfo
  {author} {\bibfnamefont {G.}~\bibnamefont {Weill}},\ }\bibfield  {title}
  {\bibinfo {title} {Persistence length of single-stranded dna},\ }\href
  {https://doi.org/10.1021/ma970381+} {\bibfield  {journal} {\bibinfo
  {journal} {Macromolecules}\ }\textbf {\bibinfo {volume} {30}},\ \bibinfo
  {pages} {5763} (\bibinfo {year} {1997})}\BibitemShut {NoStop}%
\bibitem [{\citenamefont {Murphy}\ \emph {et~al.}(2004)\citenamefont {Murphy},
  \citenamefont {Rasnik}, \citenamefont {Cheng}, \citenamefont {Lohman},\ and\
  \citenamefont {Ha}}]{Murphy:2004}%
  \BibitemOpen
  \bibfield  {author} {\bibinfo {author} {\bibfnamefont {M.~C.}\ \bibnamefont
  {Murphy}}, \bibinfo {author} {\bibfnamefont {I.}~\bibnamefont {Rasnik}},
  \bibinfo {author} {\bibfnamefont {W.}~\bibnamefont {Cheng}}, \bibinfo
  {author} {\bibfnamefont {T.~M.}\ \bibnamefont {Lohman}},\ and\ \bibinfo
  {author} {\bibfnamefont {T.}~\bibnamefont {Ha}},\ }\bibfield  {title}
  {\bibinfo {title} {Probing single-stranded dna conformational flexibility
  using fluorescence spectroscopy},\ }\bibfield  {booktitle} {\emph {\bibinfo
  {booktitle} {Biophysical Journal}},\ }\href
  {https://doi.org/10.1016/S0006-3495(04)74308-8} {\bibfield  {journal}
  {\bibinfo  {journal} {Biophysical Journal}\ }\textbf {\bibinfo {volume}
  {86}},\ \bibinfo {pages} {2530} (\bibinfo {year} {2004})}\BibitemShut
  {NoStop}%
\bibitem [{\citenamefont {Rechendorff}\ \emph {et~al.}(2009)\citenamefont
  {Rechendorff}, \citenamefont {Witz}, \citenamefont {Adamcik},\ and\
  \citenamefont {Dietler}}]{Rechendorff:2009}%
  \BibitemOpen
  \bibfield  {author} {\bibinfo {author} {\bibfnamefont {K.}~\bibnamefont
  {Rechendorff}}, \bibinfo {author} {\bibfnamefont {G.}~\bibnamefont {Witz}},
  \bibinfo {author} {\bibfnamefont {J.}~\bibnamefont {Adamcik}},\ and\ \bibinfo
  {author} {\bibfnamefont {G.}~\bibnamefont {Dietler}},\ }\bibfield  {title}
  {\bibinfo {title} {Persistence length and scaling properties of
  single-stranded dna adsorbed on modified graphite},\ }\href
  {https://doi.org/10.1063/1.3216111} {\bibfield  {journal} {\bibinfo
  {journal} {The Journal of Chemical Physics}\ }\textbf {\bibinfo {volume}
  {131}},\ \bibinfo {pages} {095103} (\bibinfo {year} {2009})}\BibitemShut
  {NoStop}%
\bibitem [{\citenamefont {Xiao}\ \emph {et~al.}(2009)\citenamefont {Xiao},
  \citenamefont {Plakos}, \citenamefont {Lou}, \citenamefont {White},
  \citenamefont {Qian}, \citenamefont {Plaxco},\ and\ \citenamefont
  {Soh}}]{Xiao:2009}%
  \BibitemOpen
  \bibfield  {author} {\bibinfo {author} {\bibfnamefont {Y.}~\bibnamefont
  {Xiao}}, \bibinfo {author} {\bibfnamefont {K.~J.~I.}\ \bibnamefont {Plakos}},
  \bibinfo {author} {\bibfnamefont {X.}~\bibnamefont {Lou}}, \bibinfo {author}
  {\bibfnamefont {R.~J.}\ \bibnamefont {White}}, \bibinfo {author}
  {\bibfnamefont {J.}~\bibnamefont {Qian}}, \bibinfo {author} {\bibfnamefont
  {K.~W.}\ \bibnamefont {Plaxco}},\ and\ \bibinfo {author} {\bibfnamefont
  {H.~T.}\ \bibnamefont {Soh}},\ }\bibfield  {title} {\bibinfo {title}
  {Fluorescence detection of single-nucleotide polymorphisms with a single,
  self-complementary, triple-stem dna probe.},\ }\href
  {https://doi.org/10.1002/anie.200900369} {\bibfield  {journal} {\bibinfo
  {journal} {Angew Chem Int Ed Engl}\ }\textbf {\bibinfo {volume} {48}},\
  \bibinfo {pages} {4354} (\bibinfo {year} {2009})}\BibitemShut {NoStop}%
\bibitem [{\citenamefont {Ouldridge}\ \emph {et~al.}(2013)\citenamefont
  {Ouldridge}, \citenamefont {Sulc}, \citenamefont {Romano}, \citenamefont
  {Doye},\ and\ \citenamefont {Louis}}]{Ouldridge:2013}%
  \BibitemOpen
  \bibfield  {author} {\bibinfo {author} {\bibfnamefont {T.}~\bibnamefont
  {Ouldridge}}, \bibinfo {author} {\bibfnamefont {P.}~\bibnamefont {Sulc}},
  \bibinfo {author} {\bibfnamefont {F.}~\bibnamefont {Romano}}, \bibinfo
  {author} {\bibfnamefont {J.}~\bibnamefont {Doye}},\ and\ \bibinfo {author}
  {\bibfnamefont {A.~A.}\ \bibnamefont {Louis}},\ }\bibfield  {title} {\bibinfo
  {title} {Dna hybridization kinetics: zippering, internal displacement and
  sequence dependence},\ }\href@noop {} {\bibfield  {journal} {\bibinfo
  {journal} {Nucl. Acids. Res.}\ }\textbf {\bibinfo {volume} {41}},\ \bibinfo
  {pages} {8886} (\bibinfo {year} {2013})}\BibitemShut {NoStop}%
\bibitem [{\citenamefont {Altan-Bonnet}\ \emph {et~al.}(2003)\citenamefont
  {Altan-Bonnet}, \citenamefont {Libchaber},\ and\ \citenamefont
  {Krichevsky}}]{Altan-Bonnet:2003}%
  \BibitemOpen
  \bibfield  {author} {\bibinfo {author} {\bibfnamefont {G.}~\bibnamefont
  {Altan-Bonnet}}, \bibinfo {author} {\bibfnamefont {A.}~\bibnamefont
  {Libchaber}},\ and\ \bibinfo {author} {\bibfnamefont {O.}~\bibnamefont
  {Krichevsky}},\ }\bibfield  {title} {\bibinfo {title} {Bubble dynamics in
  double-stranded dna},\ }\href@noop {} {\bibfield  {journal} {\bibinfo
  {journal} {Physical Review Letters}\ }\textbf {\bibinfo {volume} {90}},\
  \bibinfo {pages} {138101} (\bibinfo {year} {2003})}\BibitemShut {NoStop}%
\bibitem [{\citenamefont {Schneider}(1994)}]{Schneider:1994}%
  \BibitemOpen
  \bibfield  {author} {\bibinfo {author} {\bibfnamefont {T.}~\bibnamefont
  {Schneider}},\ }\bibfield  {title} {\bibinfo {title} {Sequence logos,
  machine/channel capacity, maxwell's demon, and molecular computers: a review
  of the theory of molecular machines},\ }\href@noop {} {\bibfield  {journal}
  {\bibinfo  {journal} {Nanotechnology}\ }\textbf {\bibinfo {volume} {5}},\
  \bibinfo {pages} {1} (\bibinfo {year} {1994})}\BibitemShut {NoStop}%
\bibitem [{\citenamefont {Chi}\ \emph {et~al.}(2012)\citenamefont {Chi},
  \citenamefont {Hannon},\ and\ \citenamefont {Darnell}}]{Chi:2012}%
  \BibitemOpen
  \bibfield  {author} {\bibinfo {author} {\bibfnamefont {S.~W.}\ \bibnamefont
  {Chi}}, \bibinfo {author} {\bibfnamefont {G.~J.}\ \bibnamefont {Hannon}},\
  and\ \bibinfo {author} {\bibfnamefont {R.~B.}\ \bibnamefont {Darnell}},\
  }\bibfield  {title} {\bibinfo {title} {An alternative mode of microrna target
  recognition},\ }\href@noop {} {\bibfield  {journal} {\bibinfo  {journal}
  {Nature Structural and Molecular Biology}\ }\textbf {\bibinfo {volume} {19}}
  (\bibinfo {year} {2012})}\BibitemShut {NoStop}%
\bibitem [{\citenamefont {Mishra}\ and\ \citenamefont
  {Humeniuk}(2013)}]{Mishra:2013}%
  \BibitemOpen
  \bibfield  {author} {\bibinfo {author} {\bibfnamefont {P.~J.}\ \bibnamefont
  {Mishra}}\ and\ \bibinfo {author} {\bibfnamefont {R.}~\bibnamefont
  {Humeniuk}},\ }\bibfield  {title} {\bibinfo {title} {Microrna
  polymorphisms},\ }\href {http://www.els.net [doi:
  10.1002/9780470015902.a0022428]} {\bibfield  {journal} {\bibinfo  {journal}
  {eLS}\ }\bibinfo {series} {eLS} (\bibinfo {year} {2013})}\BibitemShut
  {NoStop}%
\bibitem [{\citenamefont {Schoen}\ \emph {et~al.}(2009)\citenamefont {Schoen},
  \citenamefont {Krammer},\ and\ \citenamefont {Braun}}]{Schoen:2009}%
  \BibitemOpen
  \bibfield  {author} {\bibinfo {author} {\bibfnamefont {I.}~\bibnamefont
  {Schoen}}, \bibinfo {author} {\bibfnamefont {H.}~\bibnamefont {Krammer}},\
  and\ \bibinfo {author} {\bibfnamefont {D.}~\bibnamefont {Braun}},\ }\bibfield
   {title} {\bibinfo {title} {Hybridization kinetics is different inside
  cells},\ }\href@noop {} {\bibfield  {journal} {\bibinfo  {journal} {Proc.
  Natl. Acad. Sci. USA}\ }\textbf {\bibinfo {volume} {106}},\ \bibinfo {pages}
  {21649} (\bibinfo {year} {2009})}\BibitemShut {NoStop}%
\bibitem [{\citenamefont {Ellis}(2001)}]{Ellis:2001}%
  \BibitemOpen
  \bibfield  {author} {\bibinfo {author} {\bibfnamefont {R.~J.}\ \bibnamefont
  {Ellis}},\ }\bibfield  {title} {\bibinfo {title} {Macromolecular crowding:
  obvious but underappreciated},\ }\href@noop {} {\bibfield  {journal}
  {\bibinfo  {journal} {Trends Biochem. Sci.}\ }\textbf {\bibinfo {volume}
  {26}},\ \bibinfo {pages} {597} (\bibinfo {year} {2001})}\BibitemShut
  {NoStop}%
\bibitem [{\citenamefont {Weber}\ and\ \citenamefont
  {Brangwynne}(2012)}]{Weber:2012}%
  \BibitemOpen
  \bibfield  {author} {\bibinfo {author} {\bibfnamefont {S.}~\bibnamefont
  {Weber}}\ and\ \bibinfo {author} {\bibfnamefont {C.~P.}\ \bibnamefont
  {Brangwynne}},\ }\bibfield  {title} {\bibinfo {title} {Getting rna and
  protein in phase},\ }\href@noop {} {\bibfield  {journal} {\bibinfo  {journal}
  {Cell}\ }\textbf {\bibinfo {volume} {149}},\ \bibinfo {pages} {1188}
  (\bibinfo {year} {2012})}\BibitemShut {NoStop}%
\end{thebibliography}%

\end{document}